\documentclass[onecolumn,letterpaper, 10 pt,journal]{IEEEtran}

\IEEEoverridecommandlockouts
\usepackage{cite}
\usepackage{amsmath,amssymb,amsfonts}
\usepackage{amsthm}
\usepackage{algorithmic}
\usepackage{graphicx}
\usepackage{breqn}
\usepackage{booktabs}
\usepackage{textcomp}
\usepackage{xcolor}
\def\BibTeX{{\rm B\kern-.05em{\sc i\kern-.025em b}\kern-.08em
    T\kern-.1667em\lower.7ex\hbox{E}\kern-.125emX}}
\title{\LARGE \bf
	Virtual Reference Feedback Tuning for linear discrete-time systems with robust stability guarantees based on Set Membership}
\author{William D'Amico and Marcello Farina%
	\thanks{The authors are with Dipartimento di Elettronica, Informazione e Bioingegneria,
		Politecnico di Milano, Via Ponzio 24/5, 20133, Milano, Italy. \\
		{Corresponding author: \tt\small william.damico@polimi.it}}%
}
\begin{document}
%
%

\maketitle
\thispagestyle{empty}
\pagestyle{empty}

\begin{abstract}
	In this paper we propose a novel methodology that allows to design, in a purely data-based fashion and for linear single-input and single-output systems, both robustly stable and performing control systems for tracking piecewise constant reference signals.
	The approach uses both (i) Virtual Reference Feedback Tuning for enforcing suitable performances and (ii) the Set Membership framework for providing \textit{a-priori} robust stability guarantees. Indeed, an uncertainty set for the system parameters is obtained through Set Membership identification, where an algorithm based on the scenario approach is proposed to estimate the inflation parameter in a probabilistic way. Based on this set, robust stability conditions are enforced as Linear Matrix Inequality constraints within an optimization problem whose linear cost function relies on Virtual Reference Feedback Tuning.
	To show the generality and effectiveness of our approach, we apply it to two of the most widely used yet simple control schemes, i.e., where tracking is achieved thanks to (i) a static feedforward action and (ii) an integrator in closed-loop.\\ 
	The proposed method is not fully direct due to the Set Membership identification. However, the uncertainty set is used with the only objective of providing robust stability guarantees for the closed-loop system and it is not directly used for the performances optimization, which instead is totally based on data. The effectiveness of the developed method is demonstrated with reference to two simulation examples. A comparison with other data-driven methods is also carried out.
\end{abstract}


\section{Introduction}
In Automation and Control, data-based techniques are becoming increasingly popular \cite{hou2013model} since they allow to design controllers that provide satisfactory results with moderate time and computational effort. Data-based controller design methods can be divided in \textit{indirect} and \textit{direct} ones. The former methods aim at first identifying a model of the plant, based on which the controller is designed \cite{armenio2019model}. The latter methods aim at directly identifying the controller through optimization from a controller class previously selected \cite{campi2000virtual}. Notably, some algorithms recently proposed in the literature bridge the gap between these two categories, proposing hybrid approaches that exploit some the advantages of both direct and indirect methods: identification for control \cite{hjalmarsson2005experiment}, dual control \cite{feldbaum1963dual}, control-oriented identification \cite{formentin2018core}, and regularized data-enabled predictive control \cite{dorfler2022bridging}.\smallskip\\
One of the major issues in data-driven methods, especially in the \textit{direct} ones, is the possibility to provide stability guarantees of the resulting closed-loop system. Some methods to validate or certify a “data-tuned" controller before its actual implementation have been more recently proposed in the data-driven control literature, see e.g.,\cite{dehghani2009validating, cha2014verifying, da2020one}. In \cite{van2011data,battistelli2018direct,selvi2021optimal}, sufficient conditions for closed-loop stability based on small-gain arguments have been introduced in a linear setting. These conditions can be included directly in the algorithm defined for control design. Alternatively, the behavioural framework in \cite{de2019formulas} has paved the way to the inclusion of data-dependent Linear Matrix Inequalities (LMIs) for closed-loop stability in several data-driven control problems, e.g., direct data-driven linear quadratic regulator design \cite{dorfler2021certainty}, data-enabled predictive control \cite{coulson2021distributionally}, direct data-driven model-reference control \cite{breschi2021direct}, dealing also with noisy data \cite{bisoffi2021trade, de2021low}. Also, in the field of nonlinear systems, methods for robust control from data have been provided \cite{novara2014set, guo2021data}.\smallskip\\
Among \textit{direct} methods, Virtual Reference Feedback Tuning (VRFT) has gained wide popularity due to its simplicity and satisfactory performances. VRFT has been first introduced for linear controller design \cite{campi2002virtual} and has been later extended to a nonlinear setup \cite{campi2006direct}. One of the major issues in VRFT, as well as in many \textit{direct} data-driven methods, concerns the possibily of providing stability guarantees for the feedback system. In fact, based on an available batch of data, the VRFT method only allows to design a controller such that the control system is as close as possible to a given reference closed-loop model. However, especially if the controller identification results are poor and the obtained regulator is far from the ideal one, the resulting feedback system may display bad performances and even instability. However, in \cite{campi2000virtual, sala2005extensions, sala2005virtual, rojas2011internal, chiluka2021novel} controller \textit{a-posteriori} validation tests, aiming at verifying the closed-loop stability, have been proposed for linear systems specifically for VRFT.\smallskip\\
In this paper, our objective is to propose a methodology that allows to design, in a purely data-based fashion, both robustly stable and performing control systems for tracking (piecewise constant) reference signals. In particular, to show its generality and effectiveness we apply our methodology to two of the most widely used yet simple control schemes, i.e., where tracking is achieved thanks to (i) a static feedforward action and (ii) an integrator in closed-loop.\\
Stability is guaranteed during the VRFT controller design phase. To do so, we first define an uncertainty set for the system through the Set Membership (SM) identification technique (e.g., see \cite{milanese2013bounding, terzi2019learning, lauricella2020set} and the references therein). An algorithm based on the scenario approach \cite{campi2011sampling} is proposed to carry out a SM identification of the uncertainty set with probabilistic guarantees. Robust stability constraints are so enforced by means of appropriate LMIs (e.g., see \cite{de1999new, boyd1993control, kothare1996robust}). Also, a VRFT-based cost function is minimized in order to achieve the desired closed-loop performances.\\
One of the merits of this approach is that a numerically well posed LMI optimization problem allows to design the controller. Note that this approach may be classified as an hybrid one, consisting of both an indirect design part (SM) and a direct one (VRFT). Note that, however, the SM identification is conducted solely to define a set of models compatible with the data, for robust stability guarantees, and not to identify the system based on which the whole control design is carried out (which could limit the achievable performances), contrarily to classic indirect methods. On the other hand, with respect to classic VRFT, our method has the merit to provide theoretical stability conditions, by only enforcing linear constraints on the controller gain in the design phase.\\
Note that in \cite{cerone2017direct, cerone2017set} a non-iterative direct data-driven control approach which relies on SM errors-in-variables identification techniques is proposed. However, such an approach is inspired by a different direct method, i.e., the correlation based tuning framework. Also, the SM identification approach is not used to provide robust stability conditions for the closed-loop system, but to define a feasible controller parameter set.\smallskip\\

The approach is validated on two simulation examples: a minimum phase system and a non-minimum phase one. Comparisons with the classical VRFT \cite{campi2002virtual} and the approach proposed in \cite{battistelli2018direct} are also performed.\smallskip\\
The paper is organized as follows: in Section \ref{sec:PS} the control problem is defined, while in Section \ref{sec:alg} the Set Membership identification approach with probabilistic guarantees is described in details. In Section \ref{sec:cs1}, the developed algorithm is described with reference to a control scheme with feedforward action, while in Section \ref{sec:cs2} the proposed approach is shown in case a control scheme with explicit integral action is used. Also, Section \ref{sec:sim} discusses the application of the proposed algorithms to two simulation examples, while conclusions are drawn in Section \ref{sec:conclusions}.\medskip\\
\textbf{Notation}\\
We denote with $k$ the discrete-time index and with $q$ the forward shift operator (i.e., $u(k+1)=qu(k)$, for a signal $u(k)$). We indicate with $F(q)u(k)$ the signal $y(k)$ obtained by filtering the signal $u(k)$ through a discrete-time system with transfer function $F(q)$. Given a matrix $R$, the transpose is $R^T$, the transpose of the inverse is $R^{-T}$. We denote with $0_{n,m}$ a null matrix with $n$ rows and $m$ columns, and with $\mathbf{1}_n$ a column vector with all elements equal to one and of dimension $n$, whereas $I_n$ is the identity matrix of dimension $n$. Moreover, $|a|$, $a\in\mathbb{R}$, denotes the absolute value of a real number $a$, and $||F(q)||=\sqrt{\frac{1}{2\pi}\int_{-\pi}^{\pi} |F(e^{j\omega})|^2 \,d\omega}$ denotes the 2-norm of a discrete-time linear transfer function $F(q)$. Finally, we denote with $\mathbb{E}\left[X\right]$ the expected value of a random variable $X$, with $\mathcal{P}\left\{A\right\}$ the probability of an event $A$, and with $\mathcal{P}\left\{A|B\right\}$ the conditional probability of an event $A$ given an event $B$.
%
%
\section{Problem statement}
\label{sec:PS}
We consider a discrete-time linear-time-invariant (LTI) single-input and single-output (SISO) system $\mathcal{S}$ of order $n$ described by the following input-output representation:
\begin{equation}
\label{discmodel}
\centering
\left\{\begin{split}
& z(k+1)=\theta^{o^T}\phi(k) \\
& y(k)=z(k)+d(k) \\ 
\end{split}\right.
\end{equation}
where the regressor vector $\phi(k)\in\mathbb{R}^{2n}$ is defined as
\begin{align}
	\label{regressor}
	\phi(k) = 
	\big[\begin{matrix} z(k)&\dots&z(k-n+1)&u(k)& u(k-1)&\dots&u(k-n+1) \end{matrix}\big]^T
\end{align}
In \eqref{discmodel}, $u$ is the manipulable input, $z$ the ``nominal'' output, $d$ an additive measurement noise, and $y$ the measured output.
Also, $\theta^{o}=\begin{bmatrix}\theta^{o}_1&\dots&\theta^{o}_{2n}\end{bmatrix}^T\in\mathbb{R}^{2n}$ is the vector of unknown system parameters. In this paper, the problem is formulated in the SISO setting for the sake of simplicity. We make the following assumptions on the system.\smallskip\\
\textbf{Assumption 1.} \\
- The system \eqref{discmodel} is asymptotically stable; \\
- The static gain from $u$ to $z$ is different from zero; \\
- $u(k)\in\mathbb{U}\subset\mathbb{R}$ for all $k\geq-n+1$, where $\mathbb{U}$ is compact; \\
- $|d(k)|\leq\bar{d}$ for all $k\geq-n+1$, where $\bar{d}>0$ is known; \\
- The system order $n$ is known. \hfill$\square$\smallskip\\
Some remarks are in order concerning Assumption 1. Firstly, unstable plants can be addressed by means of a cascade control architecture, provided that a first stabilizing (possibly low-performing) feedback controller is available.
Also, it is worth remarking that, in case the values of the noise bound $\bar{d}$ and of the system order $n$ are not known a priori, they can be estimated from data according to the procedures proposed in \cite{lauricella2020set}. Alternatively, the system order $n$ can also be estimated from data using subspace identification methods \cite{van2012subspace}. \\
The objective of this work is to propose a novel totally data-driven control design approach that enables to devise a controller that
\begin{enumerate}
	\item provides closed-loop system robust asymptotic stability guarantees;
	\item allows to achieve perfect asymptotic tracking of constant reference signals;
	\item makes the feedback control system as similar as possible to a given reference model of interest $\mathcal{M}$ having, as requirement, an input-output delay equal to the one of the system.
\end{enumerate}
We assume that some input-output data, obtained from suitable experiments on the system $\mathcal{S}$, are available.
The main rationale behind the proposed method is based on the following steps.\\
Based on the available data, a convex uncertainty set is first learned by resorting to the SM identification approach~\cite{milanese2013bounding} as discussed in Section \ref{sec:alg}.\\
As discussed in the introduction, in this paper we will consider two most notable control configuration, showing that our approach allows to design efficient data-based controllers in both of them: (i) a state feedback one with feedforward action in Section \ref{sec:cs1} and a dynamic one endowed with an integrator in Section \ref{sec:cs2}. In both of them, the controller parameters will be computed by applying the VRFT approach~\cite{campi2002virtual}. As discussed in~\cite{campi2002virtual}, the objective of VRFT is to identify a controller such that the resulting closed-loop system is as similar as possible to a given reference closed-loop model $\mathcal{M}$. In particular, the following cost function must be minimized:
\begin{align}
\label{mrcost}
J_{MR}(\theta_C)=||(M(q)-M_{\theta_C}(q))W(q)||^2
\end{align}
where $M(q)$ is the transfer function of the reference closed-loop model, $M_{\theta_C}(q)$ is the transfer function of the adopted control system (where $\theta_C$ represents the controller free parameter vector), and $W(q)$ is a weighting function chosen by the user. As discussed in~\cite{campi2002virtual}, the cost function \eqref{mrcost} cannot be minimized since a description of the system $\mathcal{S}$, necessary to compute $M_{\theta_C}$, is not available.\\
Importantly, ad-hoc constraints based on the learned uncertainty set will be enforced on the parameter $\theta_C$ to guarantee asymptotic stability of the control scheme. This will be done robustly with respect to all possible system parameterizations, using suitable LMIs~\cite{de1999new}.
\section{Set Membership identification with probabilistic guarantees}
\label{sec:alg}
The objective of this section is to employ SM identification to compute the set of parameters, compatible with the available data, of a class of discrete-time systems of order $n$ of the type~\eqref{discmodel}. 
We refer to \cite{milanese2013bounding} and \cite{terzi2019learning} for details on the SM identification method and theory.
In section~\ref{subsec:SM} we recall the main procedure to adopt, while in the subsequent Section~\ref{subsec:SM_prob} we propose a novel method, based on the scenario approach~\cite{campi2011sampling} to inflate the size of the so-obtained parameter set in such a way to guarantee, with a prescribed probability, that the real system parameter vector belongs to it.
\subsection{Uncertainty set: Set Membership identification}
\label{subsec:SM}
First of all note that, in view of Assumption 1,~\eqref{discmodel} lies in the class of the prediction models of the type
\begin{equation}
	\label{discclass}
	y(k+1)=\theta^{T}\hat{\phi}(k)+\xi(k)+d(k+1)
\end{equation}
where 
\begin{align}
	\label{mregressor}
	\hat{\phi}(k) = 
	\begin{bmatrix}y(k)&\dots&y(k-n+1)&u(k) & u(k-1)&\dots&u(k-n+1) \end{bmatrix}^T
\end{align}
and where $\xi(k)$ essentially represents the detrimental effect of the measurement noise on the prediction capabilities of~\eqref{discclass}.\smallskip\\
%
We assume that a finite number $N_d$ of output/regressor data pairs $(y(k+1),\hat{\phi}(k))$ is available, for $k=0,\dots,N_d-1$. As a technical assumption, we consider the system parameters to lie within a compact set $\Omega\subset\mathbb{R}^{2n}$.\smallskip\\
At this point, an estimate $\underline{\lambda}$ of the prediction error upper bound is computed by solving the following linear programming (LP) optimization problem:
\begin{align}
	\label{oeb}
	\underline{\lambda}&=\min_{\theta\in\Omega,\lambda\geq0}\lambda \notag \\
	&\textrm{subject\ to} \\
	&|y(k+1)-\theta^T\hat{\phi}(k)|\leq\lambda+\bar{d}\ \ \ \ \ \ \ \ k=0,\dots,N_d-1 \notag
\end{align}
Then, it is possible to define the Feasible Parameter Set (FPS) $\Theta(\alpha)$, i.e., the set of parameter values consistent with all the prior information and the available data, as follows.
\begin{align}
	\label{fps}
	\Theta(\alpha)=\bigl\{&\theta\in\Omega:|y(k+1)-\theta^T\hat{\phi}(k)|\leq\alpha\underline{\lambda}+\bar{d},\text{ for all }k=0,\dots,N_d-1\bigr\}
\end{align}
The value $\underline{\lambda}$ is inflated by a scalar parameter $\alpha>1$ to compensate for the uncertainty caused by the use of a finite number of measurements. With a sufficiently large number of exciting data points, a practical approach is often to set the coefficient $\alpha\simeq1$. However, a theoretically sound value of $\alpha$ guaranteeing that $\theta^{o}\in\Theta(\alpha)$ with a prescribed probability, can be computed according to the novel procedure introduced in Section~\ref{subsec:SM_prob}.\smallskip\\
Since the constraints in \eqref{fps} are linear inequalities, we can compute the $n_V$ vertices $\theta^V_1$, ..., $\theta^V_{n_V}$ of the convex hull defining the FPS \cite{avis2009polyhedral}. It follows that, for all $\theta\in\Theta(\alpha)$, there exists a set of non-negative real numbers $\gamma_1$, ..., $\gamma_{n_V}$ such that $\sum_{i=1}^{n_V}\gamma_i=1$ and
\begin{equation}\label{eq:lincomb_theta}
\theta=\sum_{i=1}^{n_V}\gamma_i\theta^V_i
\end{equation}
Note that, for high-order systems, the definition of $\theta^V_1$, ..., $\theta^V_{n_V}$ may be computationally expensive. Simpler, even if more conservative, approximations of the set $\Theta(\alpha)$ can be considered, e.g., its minimum volume outer box \cite{bemporad2004inner}. 

\subsection{Computation of the inflation parameter $\alpha$ based on the scenario approach}\label{subsec:SM_prob}
In \cite{terzi2019learning}, it is proved that $\lim_{N_d\to\infty} \alpha =1^+$. However, no methods have been proposed so far to estimate $\alpha$ in the realistic case of a finite number of data. While in \cite{lauricella2020set} an invalidation test is suggested to evaluate if the chosen value of $\alpha$ is too small by checking whether the FPS is empty for a validation experiment, it is not possible to establish whether the chosen $\alpha$ is too conservative. Therefore, this section aims at providing a novel method based on the scenario approach \cite{campi2011sampling} to estimate $\alpha$ in a sound way to assess the probability and confidence that the real parameter vector $\theta^{o}$ belongs to the resultant FPS $\Theta(\alpha)$. \\
We define $\Delta=\Omega_S\times\mathbb{D}\subset\mathbb{R}^{3n+N_d}$, where $\Omega_S$ is a set of infinite cardinality containing parameters $\theta$ of LTI SISO asymptotically stable systems of order $n$ in the same representation as \eqref{discmodel} and $\mathbb{D}=[-\bar{d},\bar{d}]^{N_d+n}$. We denote with $\delta=\begin{bmatrix}
	\theta^T&\mathbf{d}^T
\end{bmatrix}^T$ any element of $\Delta$.\\
In view of Assumption 1, it is guaranteed that $\delta^{o}=\begin{bmatrix}{\theta^{o}}^T&{\mathbf{d}^{o}}^T\end{bmatrix}^T\in\Delta$, where $\mathbf{d}^{o}=\begin{bmatrix}d^{o}(-n+1)\ \dots\ d^{o}(0)\ \dots\ d^{o}(N_d)\end{bmatrix}^T\in\mathbb{R}^{n+N_d}$ is the disturbance sequence of the real experiment.
As a technical assumption, we assume that $\Delta$ is endowed with a probability distribution $\mathbb{P}_{\delta}$. In particular, we denote with $\mathbb{P}_{\theta}$ the probability distribution of $\theta$ and with $\mathbb{P}_{d}$ the probability distribution of $d(k)$, for all $k=-n+1,...,N_d$, where $\theta$ and $d(k)$ (for any $k$) are assumed uncorrelated with each other. Both $\mathbb{P}_{\theta}$ (as discussed in Section 5 of \cite{campi2000virtual} and in the references therein) and $\mathbb{P}_{d}$ can be estimated from data.
The following algorithm is proposed.
\\
\begin{tabular}{p{0.95\columnwidth}}
	\hline
	\textbf{Algorithm 1} Estimation of $\alpha$ \\
	\hline
	\begin{enumerate}
		\item Choose a violation probability $\epsilon\in(0,1)$, a confidence parameter $\beta\in(0,1)$, and a number $p$ of scenarios to be discarded.
		\item By means of the bisection algorithm, find the minimum integer $N$ solving 
		\begin{align} 						
			\label{nsceb}
			\sum_{j=0}^{p}\binom{N}{j}\epsilon^j(1-\epsilon)^{N-j}\leq\beta
		\end{align} 
		\item Generate a sample $(\delta^1,\delta^2,\dots,\delta^N)$ of $N$ independent random elements from $(\Delta,\mathbb{P}_{\delta})$, where $\delta^i=\begin{bmatrix}{\theta^i}^T&{\mathbf{d}^i}^T\end{bmatrix}^T$, $i=1,\dots,N$.
		\item For each scenario $\delta^i$, feed the fictitious system 
		\begin{equation}
			\label{fictmodel}
			\centering
			\left\{\begin{split}
				& z^{i}(k+1)=\theta^{i^T}\phi^{i}(k) \\
				& y^{i}(k)=z^{i}(k)+d^{i}(k) \\ 
			\end{split}\right.
		\end{equation}
		with the same input signal $u(k)$ used in the real experiment, starting from the same initial condition, and collect $N_d$ output/regressor data pairs.
		\item For each scenario $\delta^i$, find the minimum $\alpha^{\delta^i}$ such that the true parameter $\theta^{i}\in\Theta^{\delta^i}(\alpha^{\delta^i})$, i.e., such that $\theta^{i}$ belongs to the FPS of the scenario $\delta^i$.
		\item Discard $p$ scenarios corresponding to the ones with the greatest $\alpha^{\delta^i}$.
		\item Among the remaining $N-p$ scenarios, take the maximum $\alpha^{\delta^i}$, denoted $\alpha^*_p$. 
		\item If deemed appropriate, change $p$ and go back to step 2, otherwise terminate. 
	\end{enumerate}\\
	\hline
\end{tabular} 
\smallskip\\
The value $\alpha^*_p$ obtained through the algorithm corresponds to the estimate of $\alpha$ to be used to properly define the FPS \eqref{fps}, i.e., $\Theta(\alpha^*_p)$. The following result holds. \medskip\\
\noindent\textbf{Proposition 1.} For all $N\geq1$ fulfilling \eqref{nsceb}, it holds that $\mathcal{P}\{\theta^o\in\Theta(\alpha^*_p)\}\geq 1-\epsilon$ with probability $\geq1-\beta$.
\begin{proof}
	We consider a sample $(\delta^1,\delta^2,\dots,\delta^N)$ of $N$ independent random elements from $(\Delta,\mathbb{P}_{\delta})$, where $N$ fulfills \eqref{nsceb}. As a technical assumption, let $\alpha\in\mathbb{A}=[1,M]$, where $M$ is an arbitrarily large real number. Note that, in case of no scenario removal, $\alpha^*_0$ is the solution of the following scenario optimization program.
	\begin{align}
		&\min_{\alpha\in\mathbb{A}}\alpha 	\label{sp} \\
		&\textrm{subject\ to }
		\alpha\in\bigcap_{i=1,\dots,N}\mathbb{A}_{\delta^i} \notag
	\end{align}
	In~\eqref{sp} we set
	\begin{align}
	\label{eq:Adefs}
		\mathbb{A}_{\delta^i}=&\begin{cases}
			\mathbb{A}_{\delta^i,1} \ \ \ \text{if}\ \ \underline{\lambda}^{i}>0 \\
			\mathbb{A}_{\delta^i,2} \ \ \ \text{if}\ \ \underline{\lambda}^{i}=0\ \ \text{and}\ \ \hat{\lambda}^{i}(k)\leq0 \text{ for all } k=0,\dots,N_d-1 \\
			\mathbb{A}_{\delta^i,3} \ \ \ \text{if}\ \ \underline{\lambda}^{i}=0\ \ \text{and}\ \ \hat{\lambda}^{i}(k)>0 \text{ for at least one }\  k=0,\dots,N_d-1
		\end{cases}\end{align}
where $\mathbb{A}_{\delta^i,1}=\left\{\alpha\in\mathbb{A}:\alpha\geq\min\left(\max_{k=0,\dots,N_d-1}\hat{\lambda}^{i}(k)/\underline{\lambda}^{i}\ ,\ M\right)\right\}$, $\mathbb{A}_{\delta^i,2}=\mathbb{A}$, and $\mathbb{A}_{\delta^i,3}=\left\{M\right\}$. Also, in the previous definitions, $\hat{\lambda}^{i}(k)=|y^{i}(k+1)-\theta^{i^T}\hat{\phi}^{i}(k)|-\bar{d}$ and $y^{i}(k+1)$, $\hat{\phi}^{i}(k)$ are the data generated from $\delta^i$ according to \eqref{fictmodel}. Finally, $\underline{\lambda}^i$ is defined according to \eqref{oeb} using $y^{i}(k+1)$ and $\hat{\phi}^{i}(k)$. Note that the cost function is $\alpha$, i.e., it is linear. Moreover, $\mathbb{A}$ and $\mathbb{A}_\delta$, $\delta\in\Delta$, are convex and closed sets and the solution to \eqref{sp} obtained by discarding $p$ scenarios, denoted with $\alpha^*_p$, exists and is unique. Without loss of generality, we assume that $\alpha^*_p<M$, e.g., through the removal of a sufficient number $p$ of constraints. \\
In view of these facts, from the scenario optimization theory with constraint removal \cite{campi2011sampling}, if $N\geq1$ fulfills \eqref{nsceb}, we can state that, with probability $\geq1-\beta$, it holds that $\mathcal{P}\left\{\delta\in\Delta:\alpha^*_p\notin\mathbb{A}_{\delta}\right\}\leq\epsilon$. By recalling that $\delta^{o}\in\Delta$, the previous statement holds also setting $\delta=\delta^o$, meaning that $\mathcal{P}\left\{\alpha^*_p\notin\mathbb{A}_{\delta^o}\right\}\leq\epsilon$. We define $\hat{\lambda}(k)=|y(k+1)-\theta^{o^T}\hat{\phi}(k)|-\bar{d}$. From~\eqref{eq:Adefs}, for $M\rightarrow+\infty$, using the formula of total probability, we have
\begin{align*}
	\mathcal{P}\left\{\alpha^*_p\notin\mathbb{A}_{\delta^o}\right\}&=\mathcal{P}\left\{\alpha^*_p\notin\mathbb{A}_{\delta^o}|\ \underline{\lambda}>0\right\}\mathcal{P}\left\{\underline{\lambda}>0\right\}+\mathcal{P}\left\{\alpha^*_p\notin\mathbb{A}_{\delta^o}|\ \underline{\lambda}=0\right\}\mathcal{P}\left\{\underline{\lambda}=0\right\}= \\
	&=\mathcal{P}\left\{\alpha^*_p<\max_{k=0,\dots,N_d-1}\hat{\lambda}(k)/\underline{\lambda}\ |\ \underline{\lambda}>0\right\}\mathcal{P}\left\{\underline{\lambda}>0\right\}+\\
	&\ \ \ \ \ \ \ \ \ \ \ \ \ \ \ \ \ \ \ \ \ \ \ \ \ \ \ \ \ \ \ \ \ \ \ \ \ \ \ +\mathcal{P}\left\{\hat{\lambda}(k)>0\text{ for at least one}\  k=0,\dots,N_d-1\ |\ \underline{\lambda}=0\right\}\mathcal{P}\left\{\underline{\lambda}=0\right\}= \\
	&=\mathcal{P}\left\{\exists k=0,\dots,N_d-1 \text{ such that }\hat{\lambda}(k)>\alpha^*_p\underline{\lambda}\ |\ \underline{\lambda}>0\right\}\mathcal{P}\left\{\underline{\lambda}>0\right\}+\\&\ \ \ \ \ \ \ \ \ \ \ \ \ \ \ \ \ \ \ \ \ \ \ \ \ \ \ \ \ \ \ \ \ \ \ \ \ \ \ +\mathcal{P}\left\{\exists k=0,\dots,N_d-1 \text{ such that }\hat{\lambda}(k)>\alpha^*_p\underline{\lambda}\ |\ \underline{\lambda}=0\right\}\mathcal{P}\left\{\underline{\lambda}=0\right\}= \\
	&=\mathcal{P}\left\{\exists k=0,\dots,N_d-1 \text{ such that }\hat{\lambda}(k)>\alpha^*_p\underline{\lambda}\right\}\leq\epsilon
\end{align*} 
In view of this
$$\mathcal{P}\left\{\exists k=0,\dots,N_d-1 \text{ such that }|y(k+1)-\theta^{o^T}\hat{\phi}(k)|>\alpha^*_p\underline{\lambda}+\bar{d}\right\}\leq\epsilon$$
This is equivalent to state that
$$\mathcal{P}\left\{ |y(k+1)-\theta^{o^T}\hat{\phi}(k)|\leq \alpha^*_p\underline{\lambda}+\bar{d}\text{ for all }k=0,\dots,N_d-1\right\}\geq 1-\epsilon$$
The proof is concluded by recalling the definition~\eqref{fps} of the FPS, in view of which $|y(k+1)-\theta^{o^T}\hat{\phi}(k)|\leq \alpha^*_p\underline{\lambda}+\bar{d}$ for all $k=0,\dots,N_d-1$ is equivalent to state that $\theta^o\in\Theta(\alpha^*_p)$.
\end{proof}
Note that removing constraints from the scenario program allows to avoid too conservative estimations of the inflation parameter $\alpha$. This is particularly useful in case of scenarios in which $\underline{\lambda}^{i}=0$ and $\hat{\lambda}^{i}(k)>0$ for at least one $k=0,\dots,N_d-1$, which may occur due to the finite dataset length and if the data are not informative enough. 

\subsection{State-space representation of the system $\mathcal{S}$}
\label{subsec:SS}
The uncertain system derived from the procedure sketched in Section~\eqref{subsec:SM} can be recast in state-space as follows:
\begin{equation}
	\label{ssrep}
	\centering
	\left\{\begin{split}
		& x(k+1)=Ax(k)+Bu(k)+B_w w(k) \\
		& y(k)=Cx(k) \\ 
	\end{split}\right.
\end{equation}
where 
\begin{align}
\label{state}
x(k) = 
\big[\begin{matrix} y(k)&\dots&y(k-n+1) & u(k-1)&\dots&u(k-n+1) \end{matrix}\big]^T\in\mathbb{R}^{2n-1},
\end{align}
\begin{align*}
A &=\left[\begin{array}{cc}
\begin{array}{ccc}\theta_1&\dots&\theta_n\end{array}&\begin{array}{ccc}\theta_{n+2}&\dots&\theta_{2n}\end{array}\\
\begin{array}{cc}I_{n-1}&0_{n-1,1}\end{array}&0_{n-1,n-1}\\
0_{1,n}&0_{1,n-1}\\
0_{n-2,n}&\begin{array}{cc}I_{n-2}&0_{n-2,1}\end{array}
\end{array}\right]
\end{align*} 
and $B=\begin{bmatrix}\theta_{n+1}&0_{1,n-1}&1&0_{1,n-2}\end{bmatrix}^T$, $B_w^T=C=\begin{bmatrix}1&0_{1,2n-2}\end{bmatrix}$. Here, $w(k)=\xi(k)+d(k+1)$ depends upon the exogenous disturbance $d$. Note that matrices $A$ and $B$ are uncertain but, in view of~\eqref{eq:lincomb_theta}
\begin{equation}
\label{eq:lincombA}
\begin{bmatrix}
A&B
\end{bmatrix}=\sum_{i=1}^{n_{V}}\gamma_i\begin{bmatrix}
A_i&B_i
\end{bmatrix}\end{equation}
where $A_i$ and $B_i$, $i=1,\dots,n_V$, are defined as above based on the known parameter vectors $\theta^V_i$, $i=1,\dots,n_V$ defined based on the SM procedure sketched in Section~\ref{subsec:SM}.
\section{VRFT with robust stability guarantees: feedforward action}
\label{sec:cs1}
\subsection{The control law}
Given a (possibly time-varying) reference signal $\bar{y}(k)$, the aim of this section is to tune the uncertain parameters of the control law
\begin{align}
\label{controllerFF0}
\centering
u(k)=&\bar{u}(k)+K(x(k)-\bar{x}(k))
\end{align}
in order to achieve the goals specified in Section~\ref{sec:PS}. In~\eqref{controllerFF0}, $\bar{u}(k)$ and $\bar{x}(k)$ are computed as the steady-state input and state, respectively, corresponding to the reference $\bar{y}(k)$ at each time instant. To this respect, in this section the following assumption is made.\smallskip\\
\textbf{Assumption 2.} \\
The static gain $\mu\in\mathbb{R}$ from $u$ to $z$ of system~\eqref{discmodel} is known. \hfill$\square$\smallskip\\
Note that the latter is a mild assumption since the static gain $\mu$ can be easily estimated from data. The advantage of Assumption 2 is that it allows to compute $\bar{u}(k)$ and $\bar{x}(k)$ at each time instant as $\bar{u}(k)=\rho \bar{y}(k)$ and $\bar{x}(k) =$\break $\begin{bmatrix}\bar{y}(k)&\dots&\bar{y}(k)&\rho\bar{y}(k)&\dots&\rho\bar{y}(k)\end{bmatrix}^T=\mathbf{f}\bar{y}(k)$, where $\rho=\mu^{-1}$ and
\begin{align}
\label{state0}
\mathbf{f} = 
\begin{bmatrix}\mathbf{1}_n\\\\\rho\mathbf{1}_{n-1}\end{bmatrix}\in\mathbb{R}^{2n-1}
\end{align}
For notational compactness we write, from~\eqref{controllerFF0},
\begin{align}
\label{controllerFF}
\centering
u(k)=f_K \bar{y}(k)+Kx(k)
\end{align}
where $f_K=\rho-K\mathbf{f}$. The only tuning parameter in this case is $K^T=\begin{bmatrix}k_1&\dots&k_{2n-1}\end{bmatrix}^T\in\mathbb{R}^{2n-1}$, which will be obtained by minimizing a suitable VRFT-based cost function. Note that the feedforward additive term $f_K \bar{y}(k)$ allows to achieve a zero steady-state error.
\subsection{Controller design}
\label{sec:vrftbcfn}
In this section we discuss how to properly design the controller parameter $K$ in the control law \eqref{controllerFF0} in order to solve the problem
\begin{align}
\label{lmiVRFTMR}
&\min_{K} J_{MR}(K)
\end{align}
where $J_{MR}(K)$ corresponds with~\eqref{mrcost} with $\theta_C=K$. Note that, in our setup, the noise $d(k)$ is non-negligible and significantly affects the system output. The presence of such noise could entail a biased definition of the controller parameter, which may cause a deterioration of the closed-loop performances. Therefore, the VRFT method will be applied by resorting to the instrumental variable approach~\cite{ljung1998system} to remove the bias.
To apply the instrumental variable approach, we need two different output datasets (denoted $y^1(k)$ and $y^2(k)$) obtained by means of two experiments performed on the plant \eqref{discmodel} with the same input sequence, i.e., $u(k)$, but each affected by a different noise sequence (denoted $d^1(k)$ and $d^2(k)$, respectively), for $k=-n+1,\dots,N_d$. If an additional experiment is not possible, a second output dataset can be alternatively obtained after a plant identification phase, as suggested in~\cite{campi2002virtual} in Section 4.1. The following assumption is required.\smallskip\\
\textbf{Assumption 3.} \\
The signals $u(k)$, $d^1(k)$, and $d^2(k)$ are uncorrelated with each other, where $d^1(k)$ and $d^2(k)$ are stationary zero-mean processes.\hfill$\square$\smallskip\\
As done in~\cite{campi2000virtual}, the VRFT method requires to define a \textit{virtual reference} sequence for each experiment $i=1,2$, i.e., ${r}^i(k)=M^{-1}(q)y^i(k)$, for all $k=-n+1,\dots,N_d-1$. Moreover, as explained in~\cite{campi2000virtual}, the data need to be be filtered using an ad-hoc filter with transfer function $F(q)$, which will be defined later. The filtered data are defined as $u_F(k)=F(q)u(k)$ and, for $i=1,2$, $y_F^i(k)=F(q)y^i(k)$, ${r}^i_F(k)=F(q){r}^i(k)$. Also, we define, for $i=1,2$ and for all $k=0,\dots,N_d-1$, $x^i_F(k) = F(q)x^i(k)=
\big[\begin{matrix} y^i_F(k)&\dots&y^i_F(k-n+1) &u_F(k-1)&\dots&u_F(k-n+1) \end{matrix}\big]^T$. Based on these sequences we can define, for each $i=1,2$,
$$\mathbf{u}_{N_d}^i=\begin{bmatrix}
u_F(0)-\bar{u}^i_F(0)&\dots&u_F(N_d-1)-\bar{u}_F^i(N_d-1)
\end{bmatrix}^T$$ 
where, for all $k=0,\dots,N_d-1$, $\bar{u}_F^i(k)=F(q)\bar{u}^i(k)=\rho {r}_F^i(k)$. Also, for $i=1,2$, we need to define the matrix
$$\mathbf{x}_{N_d}^i=\begin{bmatrix}
(x^i_F(0)-\bar{x}^i_F(0))^T\\
\vdots\\
(x_F^i(N_d-1)-\bar{x}^i_F(N_d-1))^T
\end{bmatrix}$$
where, for all $k=0,\dots,N_d-1$, $\bar{x}^i_F(k)=F(q)\bar{x}^i(k)=\mathbf{f}{r}^i_F(k)$. We finally define $$R_{N_d}=\frac{1}{2N_d}\left((\mathbf{x}^1_{N_d})^T\mathbf{u}^2_{N_d}+(\mathbf{x}^2_{N_d})^T\mathbf{u}_{N_d}^1\right)$$ $$Q_{N_d}=\frac{1}{2N_d}\left((\mathbf{x}^1_{N_d})^T\mathbf{x}^2_{N_d}+(\mathbf{x}^2_{N_d})^T\mathbf{x}^1_{N_d}\right)$$
The following assumption, which is commonly verified under mild identifiability conditions, is necessary for guaranteeing the existence of a solution to the VRFT-based optimization problem.\smallskip\\
\textbf{Assumption 4.} \\
Matrix $Q_{N_d}$ is positive definite.\hfill$\square$\smallskip\\
The following theorem provides the main tool for control design.\smallskip\\
\noindent\textbf{Theorem 1.}\\
The optimization problem
\begin{align}
\label{lmiVRFTFFN}
&\min_{L,\sigma} \sigma \\
&\textrm{subject\ to} \notag \\
\label{lmiVRFTFFNLMI}
&\begin{bmatrix}
\sigma+2LQ_{N_d}^{-1}R_{N_d}-R_{N_d}^TQ_{N_d}^{-1}GQ_{N_d}^{-1}R_{N_d}&L\smallskip\\
L^T&G
\end{bmatrix}\succcurlyeq0
\end{align}
for $N_d\rightarrow+\infty$, is equivalent to~\eqref{lmiVRFTMR}
if we set, for any scalar $\gamma>0$,
\begin{align}
&|F(e^{j\omega})|^2=\frac{|M(e^{j\omega})|^2|M_K(e^{j\omega})|^2|W(e^{j\omega})|^2} {|f_K|^2\Phi_z(\omega)} \label{filtereqN}\\
&G=\gamma Q_{N_d}\label{G_equality}
\end{align}
where $K=LG^{-1}$, and $\Phi_z(\omega)$ is the spectral density of $z(k)$. \\
Moreover if, for all $i=1,\dots,n_V$, there exist symmetric matrices $P_i$ such that
\begin{align}
\label{lmiFF}
\begin{bmatrix}
P_i & A_iG+B_iL \\
(A_iG+B_iL)^T & G+G^T-P_i
\end{bmatrix}\succ0
\end{align}
then the closed-loop system is asymptotically stable for all $\theta\in\Theta(\alpha^*_p)$.\hfill$\square$\medskip\\
Some remarks are due.\\
Firstly, the results are asymptotic (i.e., they hold for $N_d\rightarrow+\infty$), but, with a sufficiently large number of exciting data, acceptable results can be achieved.\\
Secondly, the filter $F(q)$ in \eqref{filtereqN} cannot be realized since it depends on the system $\mathcal{S}$. Nevertheless, in practice, satisfactory results are obtained by approximating $M_K(q)\simeq M(q)$, and by neglecting $f_K$. Note that $f_K$ is only a scaling factor constant in the frequency domain. Note also that $\Phi_z(\omega)$ can be estimated from data, e.g., thanks to the availability of two datasets $y^1(k)$ and $y^2(k)$ generated according to independent noise sequences. 
Therefore, the following practicable approximation of the filter is proposed:
\begin{align}
\label{filterapproxN}
F(q)=&\frac{(M(q))^2W(q)}{Z(q)}
\end{align}
where $Z(q)$ is such that $|Z(e^{j\omega})|^2=\Phi_z(\omega)$.\\
Finally, setting $G$ as in \eqref{G_equality} may lead to an infeasible problem, especially in combination with the stability constraint \eqref{lmiFF}. Hence, condition \eqref{G_equality} can be relaxed by defining the matrix $G$ and the scalar $\gamma$ as free optimization variables and replacing \eqref{G_equality} with the constraints:
\begin{subequations} \label{GapproxN}
	\begin{align}
		G-\gamma Q_{N_d}+\lambda_g I_{2n-1}&\succcurlyeq0 \\
		-G+\gamma Q_{N_d}+\lambda_g I_{2n-1}&\succcurlyeq0 
	\end{align}
\end{subequations}
where the scalar $\lambda_g\geq0$ has to be minimized together with $\sigma$, through a user-defined weight $c>0$.
\subsection{The algorithm}
\label{smvrftFF}
Based on the previous results and considerations, we are now in the position to describe the proposed algorithm. The steps are the following. \smallskip\\
\\
\begin{tabular}{p{0.95\columnwidth}}
	\hline
	\textbf{Algorithm 2} FF-SM-VRFT with robust stability guarantees \\
	\hline
	\begin{enumerate}
		\item With the same input, collect two input-output datasets from the plant $u(k),y^1(k),y^2(k)$, for $k=-n+1,\dots,N_d$.
		\item Compute the vertices $\theta_i^V$, $i=1,...,n_V$, of the convex uncertainty set $\Theta(\alpha^*_p)$ and the corresponding state-space matrices $A_i$ and $B_i$, $i=1,...,n_V$, by means of Algorithm 1 and the SM identification procedure described in Section~\ref{sec:alg}.
		\item Construct the vector $R_{N_d}$ and the matrix $Q_{N_d}$ using the filter \eqref{filterapproxN}.
		\item Solve the following LMI optimization problem
		$$\!\begin{aligned}
		\label{lmi4FF}
		&\min_{G,\gamma,L,\sigma,\lambda_g,P_1,...,P_{n_V}} \sigma+c\lambda_g \\
		&\textrm{subject\ to} \notag \\
		&- \text{LMI~\eqref{lmiFF} for all }i=1,...,n_V \notag\\		
		&- \text{LMI~\eqref{lmiVRFTFFNLMI}}\notag\\		
		&- \text{LMI~\eqref{GapproxN}} \notag
		\end{aligned}$$
		where $G,P_1,...,P_{n_V}$ are symmetric matrices.
		\item If the problem is feasible, then compute 
		$$K=LG^{-1}$$
	\end{enumerate}\\
	\hline
\end{tabular} 

\subsection{Proof of Theorem 1}
The first step of the proof consists of showing that the optimization problem \eqref{lmiVRFTFFN}-\eqref{lmiVRFTFFNLMI} under \eqref{G_equality} is equivalent to minimizing a VRFT-based cost function in case the instrumental variable approach is used to cope with noise and the data are filtered by $F(q)$, i.e., \begin{align}J_{VR}^{N_d}(K)=\frac{1}{N_d}\sum_{k=0}^{N_{d}-1}\left(F(q)(u(k)-\hat{u}_K^1(k))\right)\left(F(q)(u(k)-\hat{u}_K^2(k))\right)\label{JVRFT}\end{align}
In \eqref{JVRFT}, $\hat{u}_K^i(k)=f_K\bar{y}^i(k)+Kx^i(k)$, for $i=1,2$, is the value that $u(k)$ takes in case the controller is active and it is defined using the available data sequence $(u(k),y^i(k))$ from equation~\eqref{controllerFF}. Also, we consider that, in the VRFT approach, we need to set $\bar{y}^i(k)={r}^i(k)=M^{-1}(q)y^i(k)$, being ${r}^i(k)$ the virtual reference sequence. Therefore, we compute $u(k)-\hat{u}_K^i(k)=u(k)-\bar{u}^i(k)-K(x^i(k)-\bar{x}^i(k))$. In view of this
$$\begin{array}{lcl}J_{VR}^{N_d}(K)&=&\frac{1}{N_d}(\mathbf{u}^1_{N_d}-\mathbf{x}^1_{N_d}K^T)^T(\mathbf{u}^2_{N_d}-\mathbf{x}^2_{N_d}K^T)\\
	&=&1/(2N_d)\left((\mathbf{u}^1_{N_d}-\mathbf{x}^1_{N_d}K^T)^T(\mathbf{u}^2_{N_d}-\mathbf{x}^2_{N_d}K^T)+(\mathbf{u}^2_{N_d}-\mathbf{x}^2_{N_d}K^T)^T(\mathbf{u}^1_{N_d}-\mathbf{x}^1_{N_d}K^T)\right)\\
	&=&\text{const}+KQ_{N_d}K^T-2KR_{N_d}\\
	&=&\frac{1}{\gamma}(\gamma\text{const}+K G K^T-2KGQ_{N_d}^{-1}R_{N_d})
\end{array}$$
where $\text{const}=\frac{1}{2N_d}\left((\mathbf{u}^1_{N_d})^T\mathbf{u}^2_{N_d}+(\mathbf{u}^2_{N_d})^T\mathbf{u}^1_{N_d}\right)$ is constant with respect to the optimization variable $K$ and where $G$ is assigned according to~\eqref{G_equality}. Since constant additive and strictly positive scaling terms do not take any role in the minimization of a cost function, minimizing $J_{VR}^{N_d}(K)$ is equivalent to minimizing
$$\begin{array}{lcl}\tilde{J}_{VR}^{N_d}(K)&=&(K^T-Q_{N_d}^{-1}R_{N_d})^TG(K^T-Q_{N_d}^{-1}R_{N_d})
\end{array}$$
Now we set $K=LG^{-1}$ and we use $L$ as optimization variable. We can write $\tilde{J}_{VR}^{N_d}(L)=LG^{-1}L^T-2LQ_{N_d}^{-1}R_{N_d}+R_{N_d}^TQ_{N_d}^{-1}GQ_{N_d}^{-1}R_{N_d}$. In view of this, the minimization of $\tilde{J}_{VR}^{N_d}$ can also be written through the following optimization problem:
\begin{align}
	\label{minineqn}
	\centering
	&\min_{L,\sigma} \sigma \\
	&\textrm{subject\ to} \notag \\
	&\sigma\geq LG^{-1}L^T-2LQ_{N_d}^{-1}R_{N_d}+R_{N_d}^TQ_{N_d}^{-1}GQ_{N_d}^{-1}R_{N_d} \notag
\end{align}
By resorting to the Schur complement, \eqref{minineqn} can be recast as \eqref{lmiVRFTFFN}-\eqref{lmiVRFTFFNLMI}.\smallskip\\
As a second step we show that, under the setting \eqref{filtereqN} and for $N_d\rightarrow+\infty$, minimizing the cost function $J_{VR}^{N_d}(K)$ \eqref{JVRFT} is equivalent to minimizing the model-reference criterion $J_{MR}(K)$ in~\eqref{lmiVRFTMR}. Asymptotically~\cite{campi2002virtual}, if $N_d\rightarrow+\infty$, $J_{VR}^{N_d}(K)\rightarrow \bar{J}_{VR}(K)$, where
\begin{align}\bar{J}_{VR}(K)=\mathbb{E}\left[\left(F(q)(u(k)-\hat{u}_K^1(k))\right)\left(F(q)(u(k)-\hat{u}_K^2(k))\right)\right]\label{JVRas0}\end{align}
Considering~\eqref{controllerFF}, we can write, for $i=1,2$, $\hat{u}_K^i(k)=f_K\bar{y}^i(k)+B_K(q)y^i(k)+C_K(q)u(k)$, where 
$B_K(q)=k_1+k_2q^{-1}+...+k_nq^{-n+1}$, and $C_K(q)=k_{n+1}q^{-1}+...+k_{2n-1}q^{-n+1}$. Consistently with the VRFT approach, $\bar{y}^i(k)=M^{-1}(q)y^i(k)$. In view of this, we can write, for $i=1,2$, $u(k)-\hat{u}_K^i(k)=(1-C_K(q))u(k)-(B_K(q)+f_KM^{-1}(q))y^i(k)$. From~\eqref{discmodel}, for $i=1,2$, $y^i(k)=P(q)u(k)+d^i(k)$, being $P(q)$ the unknown transfer function between $u$ and $z$. Therefore we can write, for brevity, that $u(k)-\hat{u}_K^i(k)=Q(q)u(k)+R(q)d^i(k)$, where $Q(q)=1-C_K(q)-(B_K(q)+f_KM^{-1}(q))P(q)$ and $R(q)=-(B_K(q)+f_KM^{-1}(q))$. This implies that
\begin{align}\bar{J}_{VR}(K)=\mathbb{E}\left[\left(F(q)Q(q)u(k)+F(q)R(q)d^1(k)\right)\left(F(q)Q(q)u(k)+F(q)R(q)d^2(k)\right)\right]\label{JVRas1}\end{align}
In view of Assumption 3, we can write that
\begin{align}\bar{J}_{VR}(K)=\mathbb{E}\left[\left(F(q)Q(q)u(k)\right)^2\right]\label{JVRas2}\end{align}
From \eqref{controllerFF} we can compute $M_K(q)$, i.e., the real closed-loop transfer function between $\bar{y}(k)$ and $y(k)$:
$$M_K(q)=\frac{f_KP(q)}{1-C_K(q)-B_K(q)P(q)}$$
We can therefore rewrite $Q(q)u(k)=(1-C_K(q)-B_K(q)P(q)-f_KP(q)M^{-1}(q))u(k)=f_K(M_K^{-1}(q)-M^{-1}(q))P(q)u(k)=f_K(M(q)-M_K(q))/(M_K(q)M(q))z(k)$.
Using the Parseval theorem, by dropping the argument $e^{j\omega}$, we obtain that
	\begin{align}
		\label{vrcost}
		\bar{J}_{VR}(K)=\frac{1}{2\pi}\int_{-\pi}^{\pi} |f_K|^2\frac{|M-M_K|^2}{|M|^2|M_K|^2}|F|^2\Phi_z \,d\omega
	\end{align}
	Using the definition of 2-norm of a discrete-time linear transfer function, it is possible to write~\eqref{lmiVRFTMR} as
	\begin{align}
		\label{mrcostfreq}
		J_{MR}(K)=&\frac{1}{2\pi}\int_{-\pi}^{\pi} |M-M_K|^2|W|^2 \,d\omega
	\end{align}
	It is now possible to see that \eqref{mrcostfreq} is equivalent to \eqref{vrcost} if \eqref{filtereqN} is used.\smallskip\\
	Finally, we address the stability claim. Since the stability properties of the linear system do not depend upon the exogenous signal $w(k)$ in~\eqref{ssrep} and the reference signal $\bar{y}(k)$, we discard them here by setting $w(k)=0$ and $\bar{y}(k)=0$.
	By considering~\eqref{ssrep} and~\eqref{controllerFF}, the control system dynamics is described by
	\begin{align}
		\label{clseFF}
		x(k+1)=(A+BK)x(k)
	\end{align}
	Recalling~\eqref{eq:lincombA}, the uncertain system \eqref{clseFF} is robustly stable for all $\theta\in\Theta(\alpha^*_p)$ with gain $K=LG^{-1}$, according to Theorem 3 in~\cite{de1999new} (considering a single uncertainty domain for the matrices $A$ and $B$), if there exist matrices $P_i$, $G$, and $L$ such that~\eqref{lmiFF} holds for all $i=1,\dots,n_V$.
	Therefore, $K$ stabilizes all the systems \eqref{ssrep} with $\theta\in\Theta(\alpha^*_p)$. This concludes the proof.
\section{VRFT with robust stability guarantees: explicit integral action}
\label{sec:cs2}
\subsection{The control law}
\begin{figure}[htbp]
	\centering
	\includegraphics[width=0.5\columnwidth]{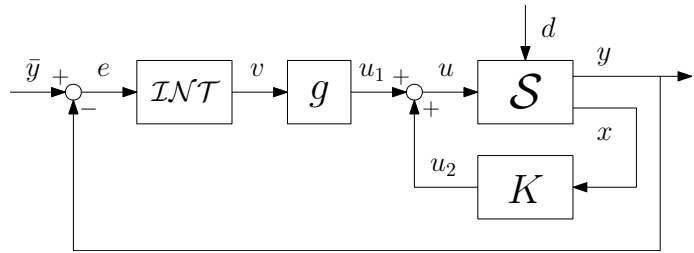}
	\caption{Control scheme}
	\label{fig: CS}
\end{figure}

In this section we propose a method for the tuning of the controller gains $K$ and $g$ in the control system depicted in Figure~\ref{fig: CS} for tracking a possibly time-varying reference signal $\bar{y}(k)$. In this scheme, an explicit integral action is introduced to achieve a zero steady-state error.
The block ``$\mathcal{INT}$'' denotes a unitary-gain integrator, with equation $v(k)=v(k-1)+e(k)$, where $e(k)=\bar{y}(k)-y(k)$. \\
The main advantage of this control scheme with respect to the one in Section~\ref{sec:cs1} is that Assumption 2 is not required, i.e., the static gain does not need to be known. Moreover, the presence of the integrator provides robustness with respect to constant disturbances and biases. \\
The controller equations can be written as
\begin{equation}
\label{controller}
\centering
\left\{\begin{split}
& \eta(k+1)=\eta(k)+e(k) \\
& u(k)=Kx(k)+g(\eta(k)+e(k)) \\ 
\end{split}\right.
\end{equation}
where $K^T=\begin{bmatrix}k_1&\dots&k_{2n-1}\end{bmatrix}^T\in\mathbb{R}^{2n-1}$ and $g\in\mathbb{R}$ are tuning parameters.
\subsection{Controller design}
\label{vrftei}
In this section we discuss how to design the controller parameters $K$ and $g$ in the controller equations~\eqref{controller} in order to solve
\begin{align}
\label{mrcostei}
\min_{K,g}J_{MR}(K,g)
\end{align}
where $J_{MR}(K,g)$ is the model reference cost function~\eqref{mrcost} with $\theta_C=\begin{bmatrix}K&g\end{bmatrix}$. As done in Section~\ref{sec:cs1}, also in this section we employ the instrumental variable approach, requiring the availability of the two datasets previously defined and consistent with Assumption~3. The corresponding virtual reference sequences are ${r}^i(k)=M^{-1}(q)y^i(k)$, for $i=1,2$ and for all $k=-n+1,\dots,N_d-1$. Therefore, the \textit{virtual error} sequences can be defined as $\tilde{e}^i(k)={r}^i(k)-y^i(k)$. After defining a filter $F(q)$ (to be later specified) and the transfer function $D(q)=1-q^{-1}$, we can define the sequences $u_{DF}(k)=D(q)F(q)u(k)$, $y^i_{DF}(k)=D(q)F(q)y^i(k)$, $\tilde{e}^i_F(k)=F(q)\tilde{e}^i(k)$, and
$$
x^i_{DF}(k) = 
\big[\begin{matrix} y^i_{DF}(k)&\dots&y^i_{DF}(k-n+1)& u_{DF}(k-1)&\dots&u_{DF}(k-n+1) \end{matrix}\big]^T
$$
We also define, for $i=1,2$,
$$\mathbf{u}_{N_d}=\begin{bmatrix}
u_{DF}(0)&\dots&u_{DF}(N_d-1)
\end{bmatrix}^T$$ 
$$\mathbf{x}^i_{N_d}=\begin{bmatrix}
x_{DF}^i(0)^T&\tilde{e}^i_F(0)\\
\vdots&\vdots\\
x_{DF}^i(N_d-1)^T&\tilde{e}_F^i(N_d-1)
\end{bmatrix}\ $$
We finally define
$$R_{N_d}=\frac{1}{2N_d}\left((\mathbf{x}^1_{N_d}+\mathbf{x}^2_{N_d})^T\mathbf{u}_{N_d}\right)$$
$$Q_{N_d}=\frac{1}{2N_d}\left((\mathbf{x}^1_{N_d})^T\mathbf{x}^2_{N_d}+(\mathbf{x}^2_{N_d})^T\mathbf{x}^1_{N_d}\right)$$
$$E=\begin{bmatrix}
	I_{2n-1}&0_{2n-1,1}\\-C&1
\end{bmatrix}$$
$$\mathcal{R}_{N_d}=E^{-1}R_{N_d}$$
$$\mathcal{Q}_{N_d}=E^{-1}Q_{N_d}E^{-T}$$
Also in this case Assumptions 3 and 4 are considered valid. The following theorem can be proved.\smallskip\\
\noindent\textbf{Theorem 2.}\\
The optimization problem
\begin{align}
\label{lmiVRFTEIN}
&\min_{L,\sigma} \sigma \\
&\textrm{subject\ to} \notag \\
\label{lmiVRFTEINLMI}
&\begin{bmatrix}
\sigma+2L\mathcal{Q}_{N_d}^{-1}\mathcal{R}_{N_d}-\mathcal{R}_{N_d}^T\mathcal{Q}_{N_d}^{-1}G\mathcal{Q}_{N_d}^{-1}\mathcal{R}_{N_d}&L\smallskip\\
L^T&G
\end{bmatrix}\succcurlyeq0
\end{align}
for $N_d\rightarrow+\infty$, is equivalent to \eqref{mrcostei} if, for any scalar $\gamma>0$, 
\begin{align}
\label{filtereqeiN}
&|F(e^{j\omega})|^2=\frac{|M(e^{j\omega})|^2|M_{Kg}(e^{j\omega})|^2|W(e^{j\omega})|^2} {|g|^2\Phi_z(\omega)} \\
\label{GequaleiN}
&G=\gamma\mathcal{Q}_{N_d}
\end{align}
where 
\begin{equation}\begin{bmatrix}K&g\end{bmatrix}=LG^{-1}E^{-1}\label{eq:Kg_INT}\end{equation} and $\Phi_z(\omega)$ is the spectral density of $z(k)$. \\
Moreover if, for all $i=1,\dots,n_V$, there exist symmetric matrices $P_i$ such that
\begin{align}
\label{lmiEI}
\begin{bmatrix}
P_i & \mathcal{A}_iG+\mathcal{B}_iL \\
(\mathcal{A}_iG+\mathcal{B}_iL)^T & G+G^T-P_i
\end{bmatrix}\succ0
\end{align}
where
$\mathcal{A}_i=\begin{bmatrix}
A_i & 0_{2n-1,1} \\
-C & 1
\end{bmatrix}$ and $\mathcal{B}_i=\begin{bmatrix}
B_i \\
0
\end{bmatrix}$, then the closed-loop system is asymptotically stable for all $\theta\in\Theta(\alpha_p^*)$.\hfill$\square$\medskip\\
With the same arguments presented in Section \ref{sec:vrftbcfn}, the practicable approximation of the filter \eqref{filterapproxN} can be used also in this case.\\
Moreover, condition \eqref{GequaleiN} can be relaxed by using both $G$ and $\gamma$ as free optimization variables and imposing the following constraints:
\begin{subequations} \label{GapproxNei}
	\begin{align}
	G-\gamma\mathcal{Q}_{N_d}+\lambda_g I_{2n}&\succcurlyeq0 \\
	-G+\gamma\mathcal{Q}_{N_d}+\lambda_g I_{2n}&\succcurlyeq0 
	\end{align}
\end{subequations}
where $\lambda_g\geq0$ must also be minimized together with $\sigma$, through a user-defined weight $c>0$.
\subsection{The algorithm}
\label{smvrftEI}
Based on the previous results and considerations, we are now in the position to describe the proposed algorithm. The steps are the following. \smallskip\\
\\
\begin{tabular}{p{0.95\columnwidth}}
	\hline
	\textbf{Algorithm 3} EI-SM-VRFT with robust stability guarantees \\
	\hline
	\begin{enumerate}
		\item With the same input, collect two input-output datasets from the plant $u(k),y^1(k),y^2(k)$, for $k=-n+1,\dots,N_d$.
		\item Compute the vertices $\theta_i^V$, $i=1,...,n_V$, of the convex uncertainty set $\Theta(\alpha^*_p)$ and the corresponding state-space matrices $\mathcal{A}_i$ and $\mathcal{B}_i$, $i=1,...,n_V$, by means of Algorithm 1 and the SM identification procedure described in Section~\ref{sec:alg}.
		\item Construct the vector $\mathcal{R}_{N_d}$ and the matrix $\mathcal{Q}_{N_d}$ using the filter \eqref{filterapproxN}.
		\item Solve the following LMI optimization problem
		$$\!\begin{aligned}
		\label{lmi4EI}
		&\min_{G,\gamma,L,\sigma,\lambda_g,P_1,...,P_{n_V}} \sigma+c\lambda_g \\
		&\textrm{subject\ to} \notag \\
		&- \text{LMI~\eqref{lmiEI} for all }i=1,...,n_V \notag\\		
		&- \text{LMI~\eqref{lmiVRFTEINLMI}}\notag\\		
		&- \text{LMI~\eqref{GapproxNei}} \notag
		\end{aligned}$$
		where $G,P_1,...,P_{n_V}$ are symmetric matrices.
		\item If the problem is feasible, then compute $K$ and $g$ according to~\eqref{eq:Kg_INT}.
	\end{enumerate}\\
	\hline
\end{tabular}
\subsection{Proof of Theorem 2}
We first show that minimizing \eqref{lmiVRFTEIN}-\eqref{lmiVRFTEINLMI} under \eqref{GequaleiN} is equivalent to minimizing a VRFT-based cost function in case the instrumental variable approach is used to cope with noise and the data are filtered by $F(q)$, i.e., \begin{align}J_{VR}^{N_d}(K,g)=\frac{1}{N_d}\sum_{k=0}^{N_{d}-1}\left(F(q)(u(k)-\hat{u}_{K,g}^1(k))\right)\left(F(q)(u(k)-\hat{u}_{K,g}^2(k))\right)\label{JVRFTEIN}\end{align}
In \eqref{JVRFTEIN}, $\hat{u}_{K,g}^i(k)$, for $i=1,2$, is the optimal predictor of $u(k)$ in case the controller is active and it is defined using the available data sequence $(u(k),y^i(k))$ according to~\eqref{controller}. Indeed, $u(k)=g(\eta(k)+e(k))+Kx(k)=\frac{g}{D(q)}e(k)+B_K(q)y(k)+C_K(q)u(k)$ where, as defined in Section~\ref{sec:cs1}, $B_K(q)=k_1+k_2q^{-1}+\dots+k_nq^{-n+1}$ and $C_K(q)=k_{n+1}q^{-1}+\dots+k_{2n-1}q^{-n+1}$. Since $D(q)(1-C_K(q))u(k)=ge(k)+D(q)B_K(q)y(k)$, it follows that the predictor is described by $\hat{u}_{K,g}^i(k)=ge^i(k)+D(q)B_K(q)y^i(k)+(1-D(q)(1-C_K(q)))u(k)$. According to the VRFT approach, we need to set $e^i(k)=\tilde{e}^i(k)$, being $\tilde{e}^i(k)$ the virtual error sequence. Therefore, we compute
\begin{equation}\begin{array}{lcl}
u(k)-\hat{u}_{K,g}^i(k)&=&D(q)(1-C_K(q))u(k)-g\tilde{e}^i(k)-D(q)B_K(q)y^i(k)\\
&=&D(q)u(k)-D(q)(B_K(q)y^i(k)+C_K(q)u(k))-g\tilde{e}^i(k)\\
&=&D(q)u(k)-D(q)Kx^i(k)-g\tilde{e}^i(k)
\end{array}\label{eq:u-uhat}\end{equation} 
Therefore, we rewrite~\eqref{JVRFTEIN} as
$$\begin{array}{lcl}J_{VR}^{N_d}(K,g)&=&\frac{1}{N_d}\sum_{k=0}^{N_{d}-1}\left(F(q)\left(D(q)u(k)-\begin{bmatrix}K&g\end{bmatrix}\begin{bmatrix}D(q)x^1(k)\\\tilde{e}^1(k)\end{bmatrix}\right)\right)\left(F(q)\left(D(q)u(k)-\begin{bmatrix}K&g\end{bmatrix}\begin{bmatrix}D(q)x^2(k)\\\tilde{e}^2(k)\end{bmatrix}\right)\right)\\
&=&\frac{1}{N_d}(\mathbf{u}_{N_d}-\mathbf{x}^1_{N_d}\begin{bmatrix}K&g\end{bmatrix}^T)^T(\mathbf{u}_{N_d}-\mathbf{x}^2_{N_d}\begin{bmatrix}K&g\end{bmatrix}^T)
\end{array}$$
Considering $L$ as optimization variable and recalling \eqref{eq:Kg_INT}, we can write
$$\begin{array}{lcl}J_{VR}^{N_d}(L)
&=&\frac{1}{N_d}(\mathbf{u}_{N_d}-\mathbf{x}^1_{N_d}E^{-T}G^{-1}L^T)^T(\mathbf{u}_{N_d}-\mathbf{x}^2_{N_d}E^{-T}G^{-1}L^T)\\
&=&1/(2N_d)((\mathbf{u}_{N_d}-\mathbf{x}^1_{N_d}E^{-T}G^{-1}L^T)^T(\mathbf{u}_{N_d}-\mathbf{x}^2_{N_d}E^{-T}G^{-1}L^T)+\\
&&\ \ \ \ \ \ \ \ \ \ \ \ \ \ \ \ \ \ \ \ \ \ \ \ \ \ \ \ \ \ \ \ \ \ \ \ \ \ \ \ \ \ \ \ \ \ \  +(\mathbf{u}_{N_d}-\mathbf{x}^2_{N_d}E^{-T}G^{-1}L^T)^T(\mathbf{u}_{N_d}-\mathbf{x}^1_{N_d}E^{-T}G^{-1}L^T))\\
&=&\text{const}
+LG^{-1}E^{-1}Q_{N_d}E^{-T}G^{-1}L^T-2LG^{-1}E^{-1}R_{N_d}
\\
&=&\text{const}
+LG^{-1}\mathcal{Q}_{N_d}G^{-1}L^T-2LG^{-1}\mathcal{R}_{N_d}
\end{array}$$
where $\text{const}=\frac{1}{N_d}(\mathbf{u}_{N_d})^T\mathbf{u}_{N_d}$ is constant with respect to the optimization variable $L$.\\
If we set $G$ according to~\eqref{GequaleiN},
$$\begin{array}{lcl}J_{VR}^{N_d}(L)
	&=&\frac{1}{\gamma}\left(\gamma\text{const}
	+LG^{-1}L^T-2L\mathcal{Q}_{N_d}^{-1}\mathcal{R}_{N_d}\right)
\end{array}$$
Since constant additive and strictly positive scaling terms do not take any role in the minimization of a cost function, minimizing $J_{VR}^{N_d}(L)$ is equivalent to minimizing
$$\begin{array}{lcl}\tilde{J}_{VR}^{N_d}(L)
	&=&(L^T-\gamma \mathcal{R}_{N_d})^TG^{-1}(L^T-\gamma \mathcal{R}_{N_d})\\
	&=&LG^{-1}L^T-2L\mathcal{Q}_{N_d}^{-1}\mathcal{R}_{N_d}+\mathcal{R}_{N_d}^T\mathcal{Q}_{N_d}^{-1}G\mathcal{Q}_{N_d}^{-1}\mathcal{R}_{N_d}
\end{array}$$
which, in turn, is equivalent to solving
\begin{align}
\label{minineqnEI}
\centering
&\min_{L,\sigma} \sigma \\
&\textrm{subject\ to} \notag \\
&\sigma\geq LG^{-1}L^T-2L\mathcal{Q}_{N_d}^{-1}\mathcal{R}_{N_d}+\mathcal{R}_{N_d}^T\mathcal{Q}_{N_d}^{-1}G\mathcal{Q}_{N_d}^{-1}\mathcal{R}_{N_d} \notag
\end{align}
By resorting to the Schur complement, \eqref{minineqnEI} can be recast as \eqref{lmiVRFTEIN}-\eqref{lmiVRFTEINLMI}.\smallskip\\
As a second step we show that, under the setting \eqref{filtereqeiN} and for $N_d\rightarrow+\infty$, minimizing the cost function $J_{VR}^{N_d}(K,g)$ \eqref{JVRFTEIN} is equivalent to minimizing $J_{MR}(K,g)$ in~\eqref{mrcostei}. Asymptotically~\cite{campi2002virtual}, if $N_d\rightarrow+\infty$, $J_{VR}^{N_d}(K,g)\rightarrow \bar{J}_{VR}(K,g)$, where
\begin{align}\bar{J}_{VR}(K,g)=\mathbb{E}\left[\left(F(q)(u(k)-\hat{u}_{K,g}^1(k))\right)\left(F(q)(u(k)-\hat{u}_{K,g}^2(k))\right)\right]\label{JVRas0IN}
\end{align}
Recall that $\tilde{e}^i(k)={r}^i(k)-y^i(k)$, ${r}^i(k)$ is obtained as ${r}^i(k)=M^{-1}(q)y^i(k)$ in view of VRFT and, from the definition of $x^i(k)$, $Kx^i(k)=B_K(q)y^i(k)+C_K(q)u(k)$. Also, $y^i(k)=P(q)u(k)+d^i(k)$, where $P(q)$ is the unknown transfer function between $u$ and $z$ in~\eqref{discmodel}.
In view of these facts, we can rewrite, for $i=1,2$, equation~\eqref{eq:u-uhat} as
$u(k)-\hat{u}_{K,g}^{i}(k)=Q(q)u(k)+R(q)d^i(k)$, where $$Q(q)=D(q)\left(1+g\frac{P(q)}{D(q)}-C_K(q)-B_{K}(q)P(q)-M^{-1}(q)g\frac{P(q)}{D(q)}\right)$$
$$R(q)=-D(q)B_K(q)-g(M^{-1}(q)-1)$$
This implies that
\begin{align}\bar{J}_{VR}(K,g)=\mathbb{E}\left[\left(F(q)Q(q)u(k)+F(q)R(q)d^1(k)\right)\left(F(q)Q(q)u(k)+F(q)R(q)d^2(k)\right)\right]\label{JVRas1IN}\end{align}
In view of Assumption 3, we can write that
\begin{align}\bar{J}_{VR}(K,g)=\mathbb{E}\left[\left(F(q)Q(q)u(k)\right)^2\right]\label{JVRas2IN}
\end{align}
From \eqref{controller} we can compute $M_{K,g}(q)$, i.e., the real closed-loop transfer function between $\bar{y}(k)$ and $y(k)$:
$$M_{K,g}(q)=\frac{g\frac{P(q)}{D(q)}}{1+g\frac{P(q)}{D(q)}-C_K(q)-B_{K}(q)P(q)}$$
We can therefore rewrite $Q(q)u(k)=D(q)(g\frac{P(q)}{D(q)}M_{K,g}^{-1}(q)-g\frac{P(q)}{D(q)}M^{-1}(q))u(k)=(M_{K,g}^{-1}(q)-M^{-1}(q))gP(q)u(k)=g(M(q)-M_{K,g}(q))/(M_{K,g}(q)M(q))z(k)$.
Using the Parseval theorem, by dropping the argument $e^{j\omega}$, we obtain that
\begin{align}
\label{vrcostIE}
\bar{J}_{VR}(K,g)=\frac{1}{2\pi}\int_{-\pi}^{\pi} |g|^2\frac{|M-M_{K,g}|^2}{|M|^2|M_{K,g}|^2}|F|^2\Phi_z \,d\omega
\end{align}
Using the definition of 2-norm of a discrete-time linear transfer function, it is possible to write \eqref{mrcostei} as
\begin{align}
\label{mrcostfreqIE}
J_{MR}(K,g)=&\frac{1}{2\pi}\int_{-\pi}^{\pi} |M-M_{K,g}|^2|W|^2 \,d\omega
\end{align}
It is now possible to see that \eqref{mrcostfreqIE} is equivalent to \eqref{vrcostIE} if \eqref{filtereqeiN} is used.\smallskip\\
Finally, we address the stability claim. Since the stability properties of the linear system do not depend upon the exogenous signal $w(k)$ in~\eqref{ssrep} and the reference signal $\bar{y}(k)$, we discard them here by setting $w(k)=0$ and $\bar{y}(k)=0$. In view of this, $e(k)=-y(k)$.
By considering~\eqref{ssrep} and~\eqref{controller} at the same time, the state of the control system is $\zeta(k)=\begin{bmatrix}x(k)^T&\eta(k)\end{bmatrix}^T$, whose dynamics is described by
\begin{align}
	\label{clse}
	\zeta(k+1)=\mathcal{A}_{CL}\zeta(k)
\end{align}
where 
$$\mathcal{A}_{CL}=
\begin{bmatrix}
	A+BK-gBC & gB \\
	-C & 1
\end{bmatrix}$$
Note that we can 
write $\mathcal{A}_{CL}=\mathcal{A}+\mathcal{B}J$ where, from~\eqref{eq:lincombA}
\begin{equation}
	\label{eq:lincombGH}
	\begin{bmatrix}
		\mathcal{A}&\mathcal{B}
	\end{bmatrix}=\sum_{i=1}^{n_{V}}\gamma_i\begin{bmatrix}
		\mathcal{A}_i&\mathcal{B}_i
\end{bmatrix}\end{equation}
and 
\begin{equation}J=\begin{bmatrix}K-gC&g\end{bmatrix}\label{eq:Jdef}\end{equation} 
takes the role of the control gain. 
The stability claim follows straightforwardly from \eqref{clse} and from the application of Theorem 3 in~\cite{de1999new}, by considering a single uncertainty domain for the matrices $\mathcal{A}$ and $\mathcal{B}$. Note that, as a solution to the LMI~\eqref{lmiEI}, the stabilizing gain is $J=LG^{-1}$. Note also that, from~\eqref{eq:Jdef},
\begin{align}
	\label{kgje}
	\begin{bmatrix}
		K&g
	\end{bmatrix}=JE^{-1}
\end{align}
and then we obtain~\eqref{eq:Kg_INT}. This concludes the proof.
\section{Simulation results}
\label{sec:sim}
The algorithms proposed in this paper are validated on two simulation examples displaying different features. 

\subsection{Example 1: Minimum phase plant}
\label{mpp}
\begin{figure}[h!]
	\centering
	\includegraphics[width=0.75\columnwidth]{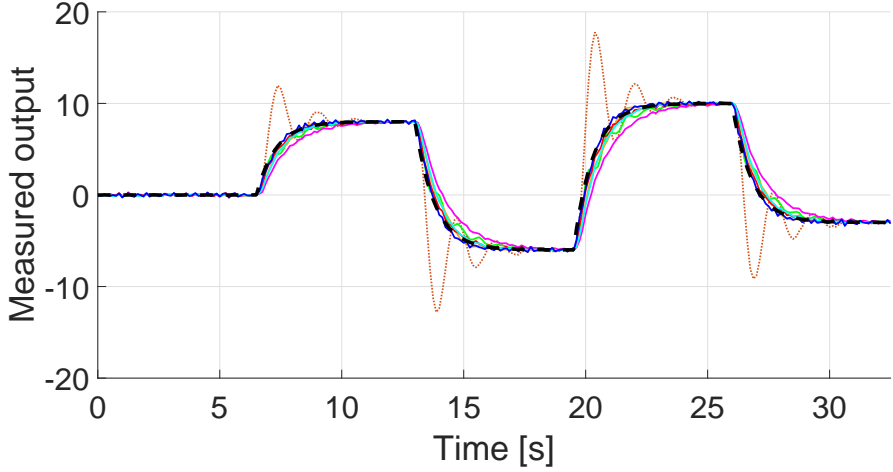}
	\caption{Measured output trajectories obtained with $\mathcal{M}_{30}$. Black dashed line: reference closed-loop trajectory; orange dotted line: output response in open-loop; blue line: FF-SM-VRFT; cyan line: FF-SM-VRFT-NF; red line: EI-SM-VRFT; magenta line: EI-SM-VRFT-NF; golden line: PID-VRFT; green line: UF.}
	\label{out30}
\end{figure}

\begin{figure}[h!]
	\centering
	\includegraphics[width=0.75\columnwidth]{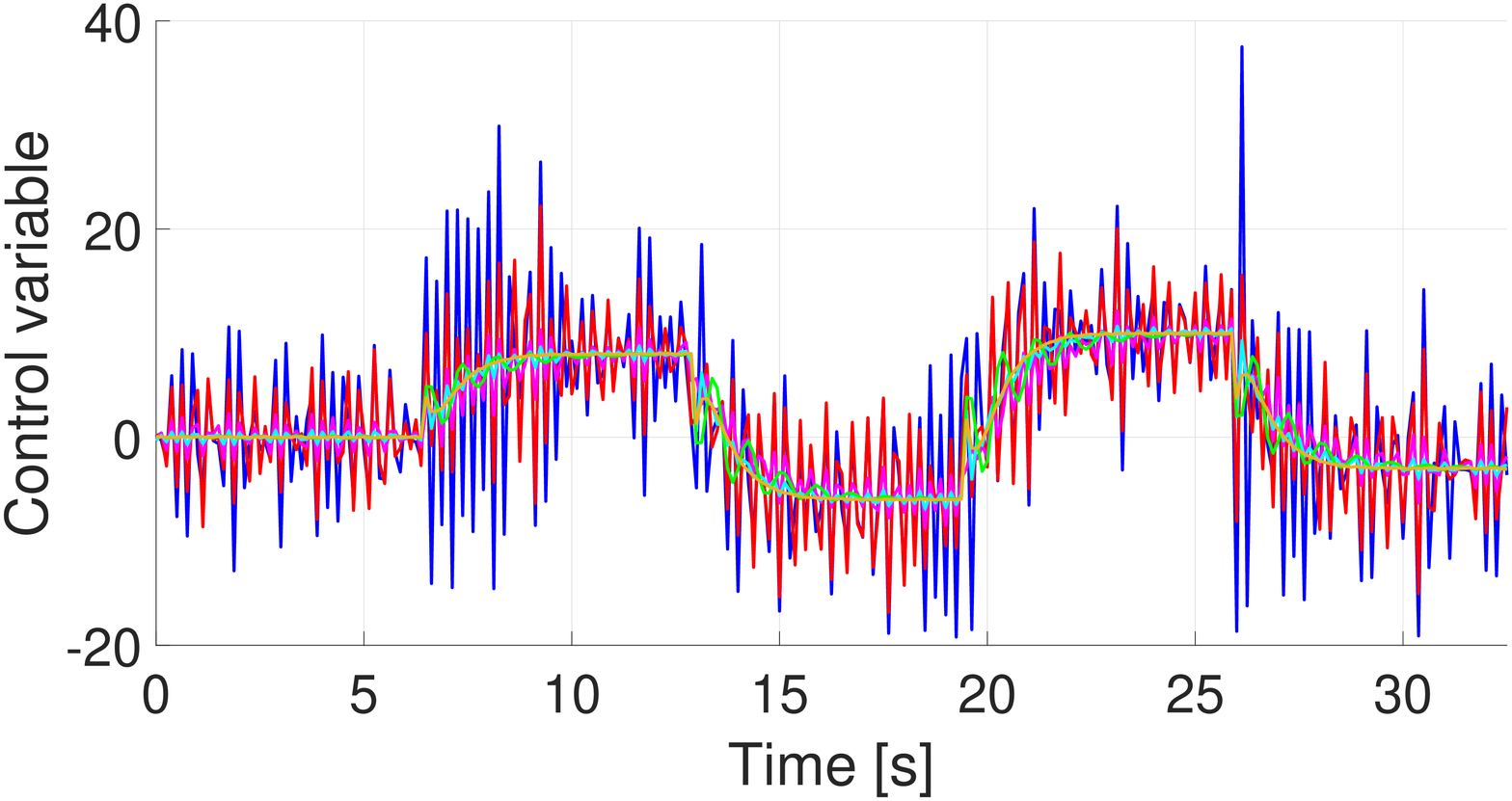}
	\caption{Input trajectories obtained with $\mathcal{M}_{30}$. Blue line: FF-SM-VRFT; cyan line: FF-SM-VRFT-NF; red line: EI-SM-VRFT; magenta line: EI-SM-VRFT-NF; golden line: PID-VRFT; green line: UF.}
	\label{cvar30}
\end{figure}

\begin{figure}[h!]
	\centering
	\includegraphics[width=0.75\columnwidth]{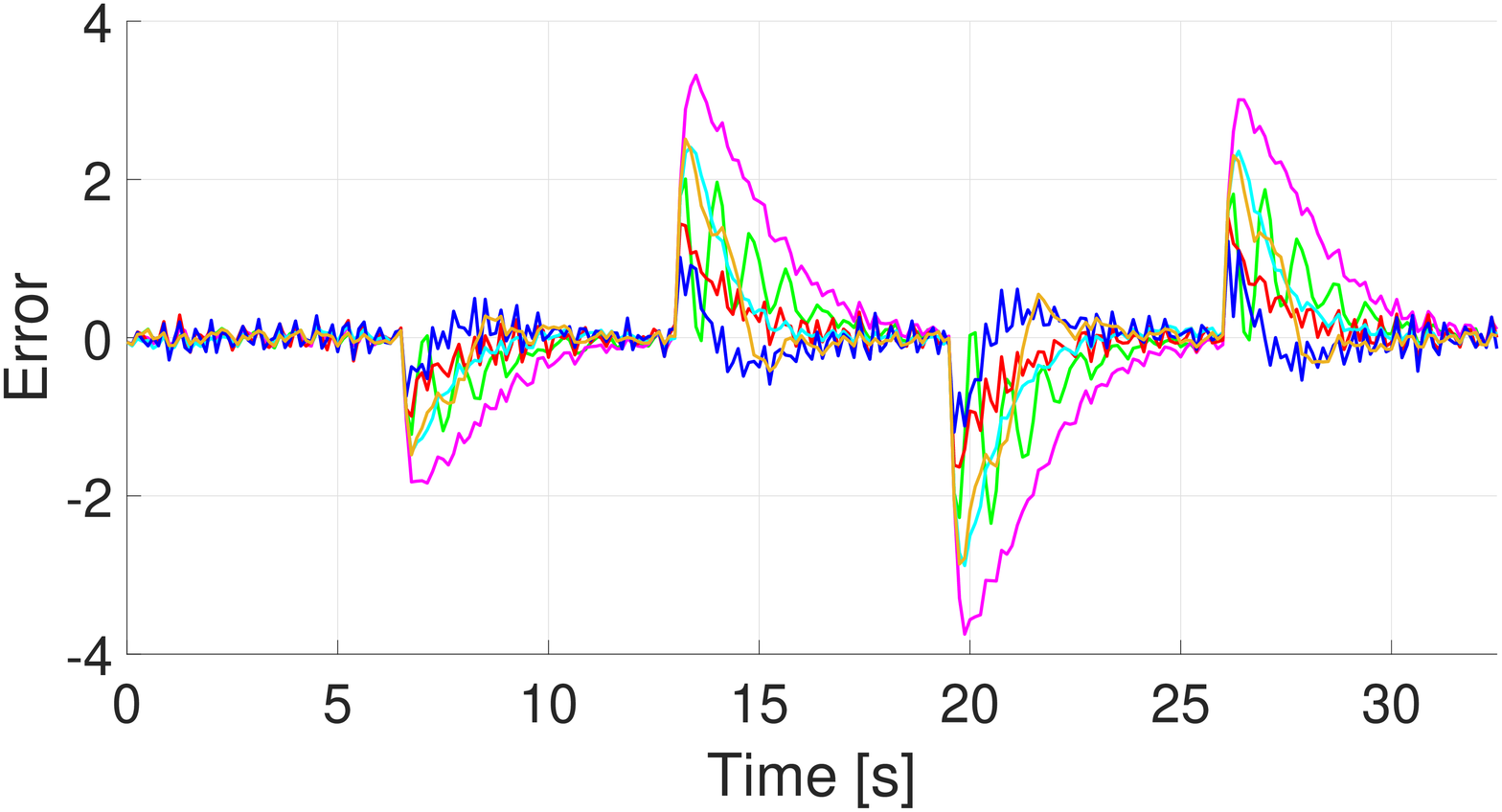}
	\caption{Error trajectories obtained with $\mathcal{M}_{30}$. Blue line: FF-SM-VRFT; cyan line: FF-SM-VRFT-NF; red line: EI-SM-VRFT; magenta line: EI-SM-VRFT-NF; golden line: PID-VRFT; green line: UF.}
	\label{err30}
\end{figure}

The considered system, drawn from~\cite{terzi2019learning}, corresponds to the discretization of the asymptotically stable system with continuous-time transfer function:
\begin{align}
	\label{sysol}
	&P(s)=\frac{Z(s)}{U(s)}=\frac{160}{(s+10)(s^2+1.6s+16)}&
\end{align}
characterized by a unitary gain and dominant complex poles with natural frequency $\omega_n=4$ and damping factor $\xi=0.2$. \\
\\
A sample time $T_s=0.125$ s is chosen. The settling time of the open-loop system is around $50T_s=6.25$ s. The system is discretized by means of the zero-order hold (ZOH) method: the corresponding nominal parameter vector is $\theta^{o}=\begin{bmatrix}1.883&-1.276&0.2346&0.0367&0.1038&0.0179\end{bmatrix}^T$ and the order is $n=3$.
An additive uniform random noise $d$ acting in the range $[-0.1,0.1]$ (i.e., $\bar{d}=0.1$) affects the output $z$ of the system.\\
Two datasets composed of $N_d=10000$ output/regressor data pairs $(y(k+1),\hat{\phi}(k))$ are collected from the plant in open-loop, with different noise realizations. The input signal is a PseudoRandom Binary Sequence (PRBS) in the range $[-10,10]$. \\
For the application of the SM method described in Section \ref{sec:alg}, to estimate the conservative factor $\alpha^*_p$ accounting for the finite dataset employed, Algorithm 1 is applied. $\mathbb{P}_{\theta}$ is estimated according to \cite{campi2000virtual} Section 5, while $\mathbb{P}_{d}$ is considered uniform in the range $[-0.1,0.1]$. We chose a violation parameter $\epsilon=0.05$, a confidence parameter $\beta=10^{-10}$, and the number of scenarios to be discarded $p=20$. Hence, the number of required scenarios obtained applying the bisection algorithm to \eqref{nsceb} is $N=1265$. As a result, we obtained $\alpha^*_p=1.1218$. The corresponding set $\Theta(\alpha^*_p)$ has $1406$ vertices. 
\begin{figure}[h!]
	\centering
	\includegraphics[width=0.75\columnwidth]{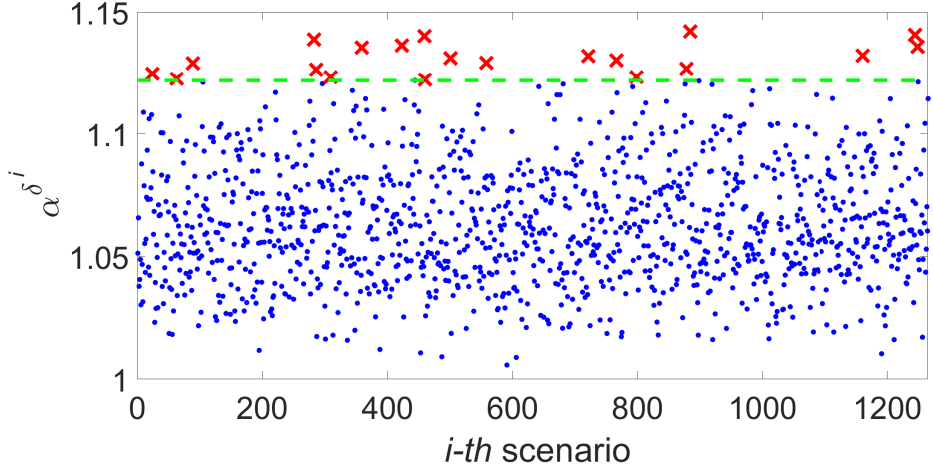}
	\caption{Scenario application: scenarios (blue dots), removed scenarios (red crosses), $\alpha^*_p$ value (green dashed line).}
	\label{sce1}
\end{figure}
In Figure \ref{sce1} the values $\alpha^{\delta^i}$ computed for each scenario $i$ are depicted. 
\begin{figure}[h!]
	\centering
	\includegraphics[width=0.75\columnwidth]{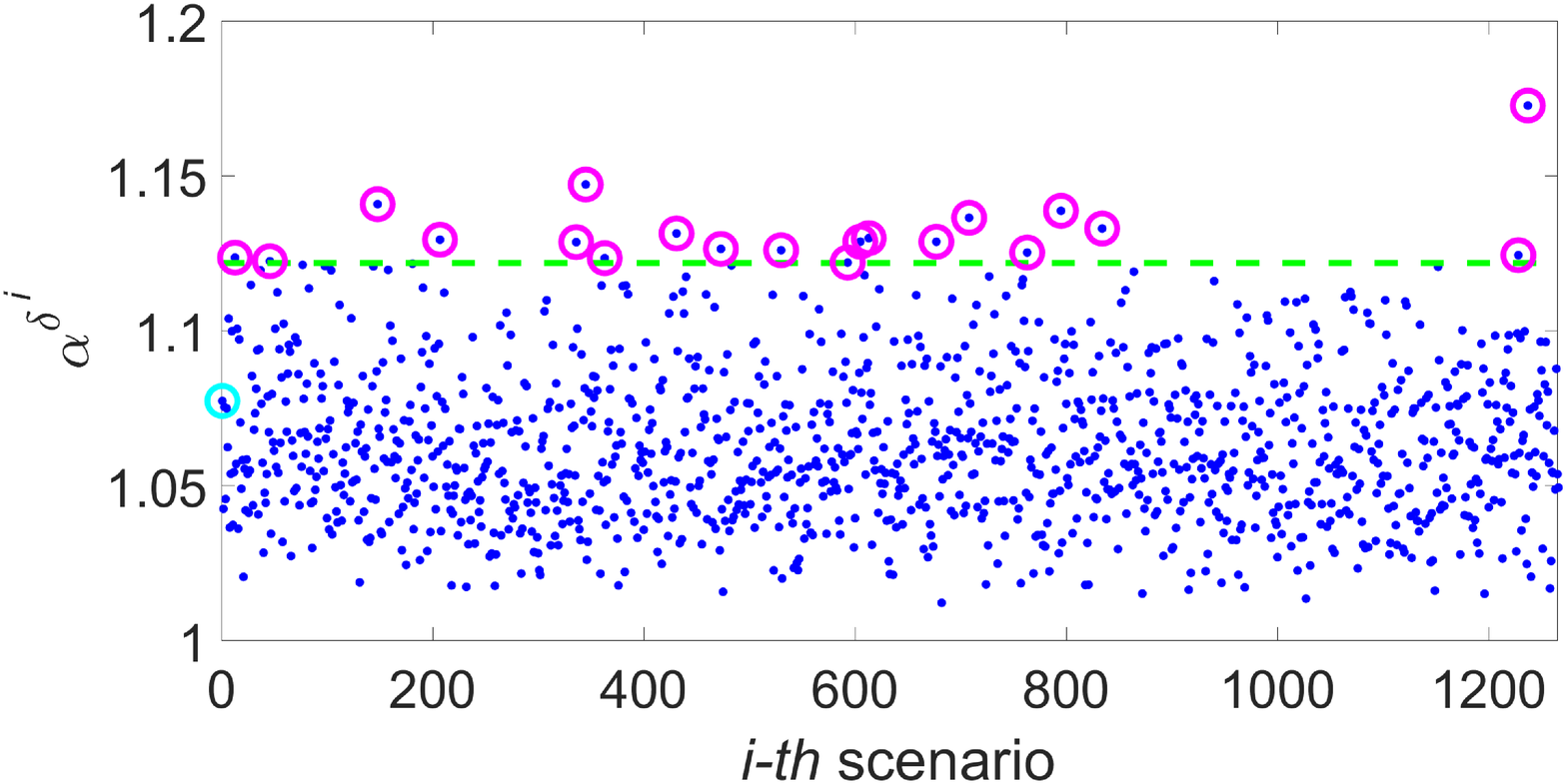}
	\caption{Validation: scenarios (blue dots), nominal case (cyan circle), violated scenarios (purple circles), $\alpha^*_p$ value (green dashed line).}
	\label{sce2}
\end{figure}
In Figure \ref{sce2} a validation test with new scenarios is carried out. The percentage of scenarios which violate $\alpha^*_p$ is $1.581\%$, lower than $\epsilon_{\%}=5\%$. Note that the nominal parameter vector $\theta^o\in\Theta(\alpha^*_p)$. \\
\\
In the following, control design tests are conducted with two different reference models $\mathcal{M}$, both characterized by a first-order asymptotically stable and unitary-gain system with equation $y_r(k) = -a_1 y_r(k-1) + b_1 r(k-1)$. The first one is denoted $\mathcal{M}_{30}$ and has settling time $30T_s=3.75$ s, being $a_1=-0.855$ and $b_1=0.145$. On the other hand, the second one, denoted $\mathcal{M}_{10}$, has settling time $10T_s=1.25$ s, being $a_1=-0.6$ and $b_1=0.4$. YALMIP and the MOSEK solver~\cite{lofberg2004yalmip,mosek} are used to solve the LMI optimization problems in Algorithms 2 and 3. \\
\\
In the simulations, the reference signal described in Table~\ref{tab: Refsignval} is used as a setpoint to evaluate the performances of the control system.\\
\begin{table}[h!]
	\centering
	\renewcommand\arraystretch{1.2}
	\caption{Reference signal values} 
	\begin{tabular}{ c c r | }
		\hline
		\multicolumn{1}{c}{$Reference$} &
		\multicolumn{1}{c}{$Interval\ (s) $} \\
		\hline
		$0$ & $[0,6.5)$ \\
		$8$ & $[6.5,13)$ \\
		$-6$ & $[13,19.5)$ \\
		$10$ & $[19.5,26)$ \\
		$-3$ & $[26,32.5]$ \\
		\hline
	\end{tabular} 
	\label{tab: Refsignval} 
\end{table}
We considered different cases for a better comparison. Namely, we tested Algorithm 2 with (FF-SM-VRFT) and without (FF-SM-VRFT-NF) the application of the filter \eqref{filterapproxN}. We also applied Algorithm 3 with (EI-SM-VRFT) and without (EI-SM-VRFT-NF) the application of the filter \eqref{filterapproxN}. We selected $c=10^6$ and $W(q)=1$. For the filter application, an estimate of $Z(q)$ was obtained from the output signal of one experiment through the identification of a discrete-time AR model of order 5. \\
Moreover, the proposed algorithms are compared with a standard VRFT linear implementation (PID-VRFT). In particular, a PID was tuned by means of the VRFT Toolbox \cite{care2019toolbox}, where the optimal filter and the IV method were applied as well. \\
Finally, a comparison is also carried out with the direct method based on controller unfalsification (UF) proposed in \cite{battistelli2018direct}, where a controller tuning procedure incorporating simple stability tests is suggested. Note that the method in \cite{battistelli2018direct} was extended in \cite{selvi2021optimal} to take into account the optimal choice of the reference models through a non-convex optimization problem. However, to have a fair comparison, we used the same reference complementary sensitivity functions considered in Algorithms 2 and 3 and in VRFT. Nevertheless, to apply the method, the choice of a suitable reference control sensitivity function $\mathcal{Q}(q)$ is also required. 
\begin{table}[h!]
	\centering
	\renewcommand\arraystretch{1.2}
	\caption{UF design parameters} 
	\begin{tabular}{ c c r | }
		\hline
		\multicolumn{1}{c}{$Case$} &
		\multicolumn{1}{c}{$Input\ sensitivity$} \\
		\hline 
		\rule{0pt}{3ex}
		$\mathcal{M}_{10}$ & $\mathcal{Q}(q)=\frac{1.5q^2-2.37q+1.23}{q^2-0.8q+0.16}$ \smallskip\\
		$\mathcal{M}_{30}$ & $\mathcal{Q}(q)=\frac{0.2181q^2-0.3483q+0.1786}{q^2-1.56q+0.6084}$ \smallskip\\
		\hline
	\end{tabular} 
	\label{tab: dparam} 
\end{table}
In Table \ref{tab: dparam} the choice of the control sensitity functions is reported\footnote{In this work, we selected $\mathcal{Q}(q)$ for UF in such a way that $\mathcal{Q}(q)P(q)$ is as similar as possible to $M(q)$, and such that the response time of the control signal is similar to the one of the desired output. Note that, however, this is the ideal tuning, which in practice cannot be done because $P(q)$ is unknown. As suggested in \cite{battistelli2018direct}, this can be approximatively done by relying on the estimated static gain and the empirical transfer function estimate. In this paper, however, we have decided to directly use $P(q)$ for showing the best performances obtainable thanks to UF and for a fair comparison.}. Furthermore, the following class of controllers is considered:
\begin{align}
	\label{ufcontr}
	C(q,\kappa)=&\frac{\kappa_1q^3+\kappa_2q^2+\kappa_3q+\kappa_4}{(q-1)(q^2+\kappa_5q+\kappa_6)}
\end{align}
where $\kappa=\begin{bmatrix}\kappa_1&\dots&\kappa_6\end{bmatrix}^T$ is the vector of tuning parameters. Accordingly, the maximum value of the weight (denoted with $\delta$ in \cite{battistelli2018direct}) for which the stability test is passed is equal to $0.95$ in case of $\mathcal{M}_{10}$, and to $0.8$ in case of $\mathcal{M}_{30}$. \smallskip\\
\begin{table}[h!]
	\centering
	\renewcommand\arraystretch{1.2}
	\caption{Spectral radius of the closed-loop system and $FIT$} 
	\begin{tabular}{ c c c r | }
		\hline
		\multicolumn{1}{c}{$Case$} &
		\multicolumn{1}{c}{$\rho$} &
		\multicolumn{1}{c}{$FIT\ (\%) $} \\
		\hline
		FF-SM-VRFT-NF $\mathcal{M}_{10}$ & $0.6749$ & $93.1148$ \\
		FF-SM-VRFT $\mathcal{M}_{10}$ & $0.8492$ & $93.5919$ \\
		EI-SM-VRFT-NF $\mathcal{M}_{10}$ & $0.6559$ & $87.8436$ \\
		EI-SM-VRFT $\mathcal{M}_{10}$ & $0.7790$ & $86.2849$ \\		
		PID-VRFT $\mathcal{M}_{10}$ & $0.8446$ & $71.4590$ \\
		UF $\mathcal{M}_{10}$ & $0.9090$ & $89.1269$ \\		
		FF-SM-VRFT-NF $\mathcal{M}_{30}$ & $0.8541$ & $87.7802$ \\
		FF-SM-VRFT $\mathcal{M}_{30}$ & $0.8214$ & $95.0641$ \\
		EI-SM-VRFT-NF $\mathcal{M}_{30}$ & $0.8970$ & $78.1134$ \\
		EI-SM-VRFT $\mathcal{M}_{30}$ & $0.8605$ & $93.1417$ \\		
		PID-VRFT $\mathcal{M}_{30}$ & $0.8913$ & $88.2519$ \\
		UF $\mathcal{M}_{30}$ & $0.9106$ & $89.4643$ \\		
		\hline
	\end{tabular} 
	\label{tab: sprad} 
\end{table}
Table \ref{tab: sprad} displays the spectral radius $\rho$ of the closed-loop system. Moreover, to test the performances in closed-loop, the following fitting index is calculated
\begin{equation}
	\label{fitsim}
	FIT (\%) = 100\cdot\left(1-\frac{\|\mathbf{y}_r-\mathbf{y}\|}{\|\mathbf{y}_r-\mathbf{\bar{y}}_r\|}\right)\in(-\infty,100]
\end{equation}
where $\mathbf{y}_r$ is the reference model output sequence, $\mathbf{y}$ is the measured output sequence and $\mathbf{\bar{y}}_r$ is a vector with all the elements equal to the mean value of the reference model output sequence $\mathbf{y}_r$.\smallskip\\
For performance evaluation, in Figures~\ref{out30} and~\ref{cvar30} we show the reference tracking results obtained with reference model $\mathcal{M}_{30}$, in terms of trajectories of the measured outputs $y(k)$ and of the control inputs $u(k)$, respectively. In Figure~\ref{err30} the output error trajectories (with respect to the reference model one) are depicted. \\
By inspection of Figures~\ref{out30} and~\ref{err30} it possible to appreciate the excellent results obtained using the methods proposed in this paper. Asymptotically stable closed-loop systems and satisfactory tracking results are achieved also with the standard VRFT linear implementation and the controller unfalsification method. Nevertheless, in general, the standard VRFT implementation does not provide any a priori closed-loop stability guarantee. Finally, note that, even if the filter \eqref{filterapproxN} is not applied, satisfactory results are obtained. 
\begin{figure}[h!]
	\centering
	\includegraphics[width=0.75\columnwidth]{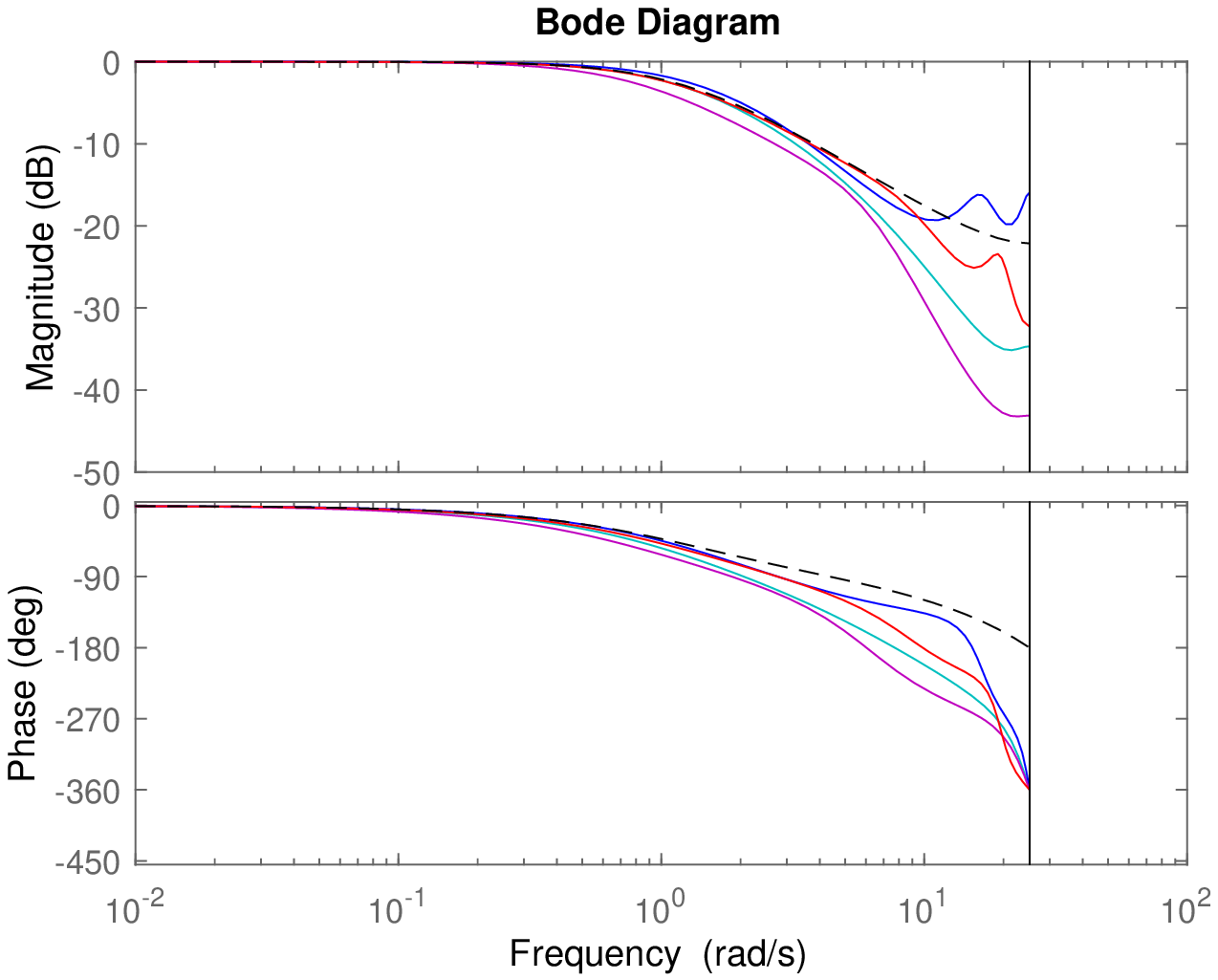}
	\caption{Bode diagrams obtained with $\mathcal{M}_{30}$. Black dashed line: Bode diagram of the reference closed-loop model; blue line: FF-SM-VRFT; cyan line: FF-SM-VRFT-NF; red line: EI-SM-VRFT; magenta line: EI-SM-VRFT-NF.}
	\label{bode30}
\end{figure}
However, the enhancing effect provided by the proposed filter is clear from Figure \ref{bode30}, where the Bode diagram of the closed-loop transfer function in the different cases is compared with the one of the reference model. \smallskip\\
\begin{figure}[h!]
	\centering
	\includegraphics[width=0.75\columnwidth]{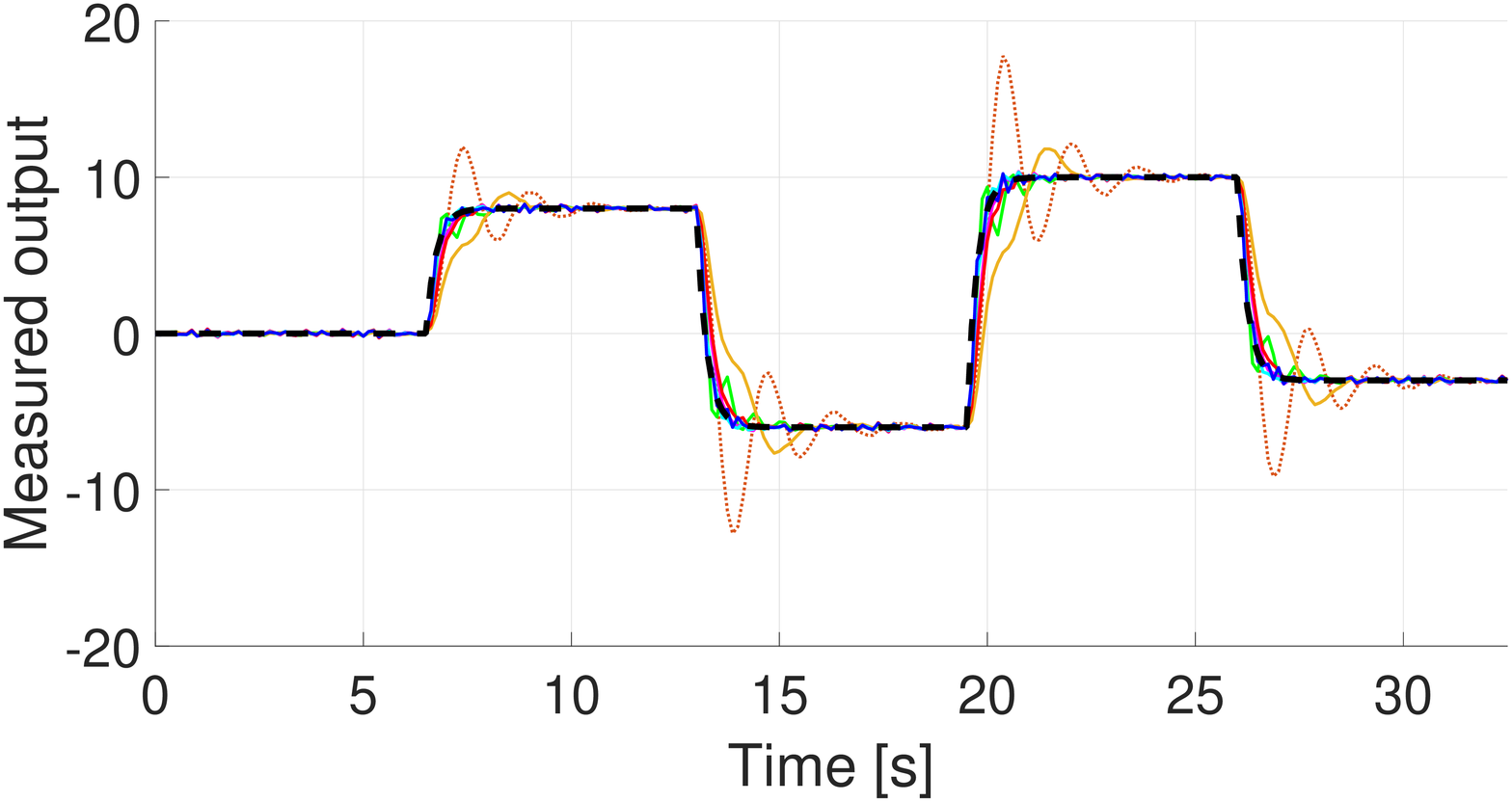}
	\caption{Measured output trajectories obtained with $\mathcal{M}_{10}$. Black dashed line: reference closed-loop trajectory; orange dotted line: output response in open-loop; blue line: FF-SM-VRFT; cyan line: FF-SM-VRFT-NF; red line: EI-SM-VRFT; magenta line: EI-SM-VRFT-NF; golden line: PID-VRFT; green line: UF.}
	\label{out10}
\end{figure}
\begin{figure}[h!]
	\centering
	\includegraphics[width=0.75\columnwidth]{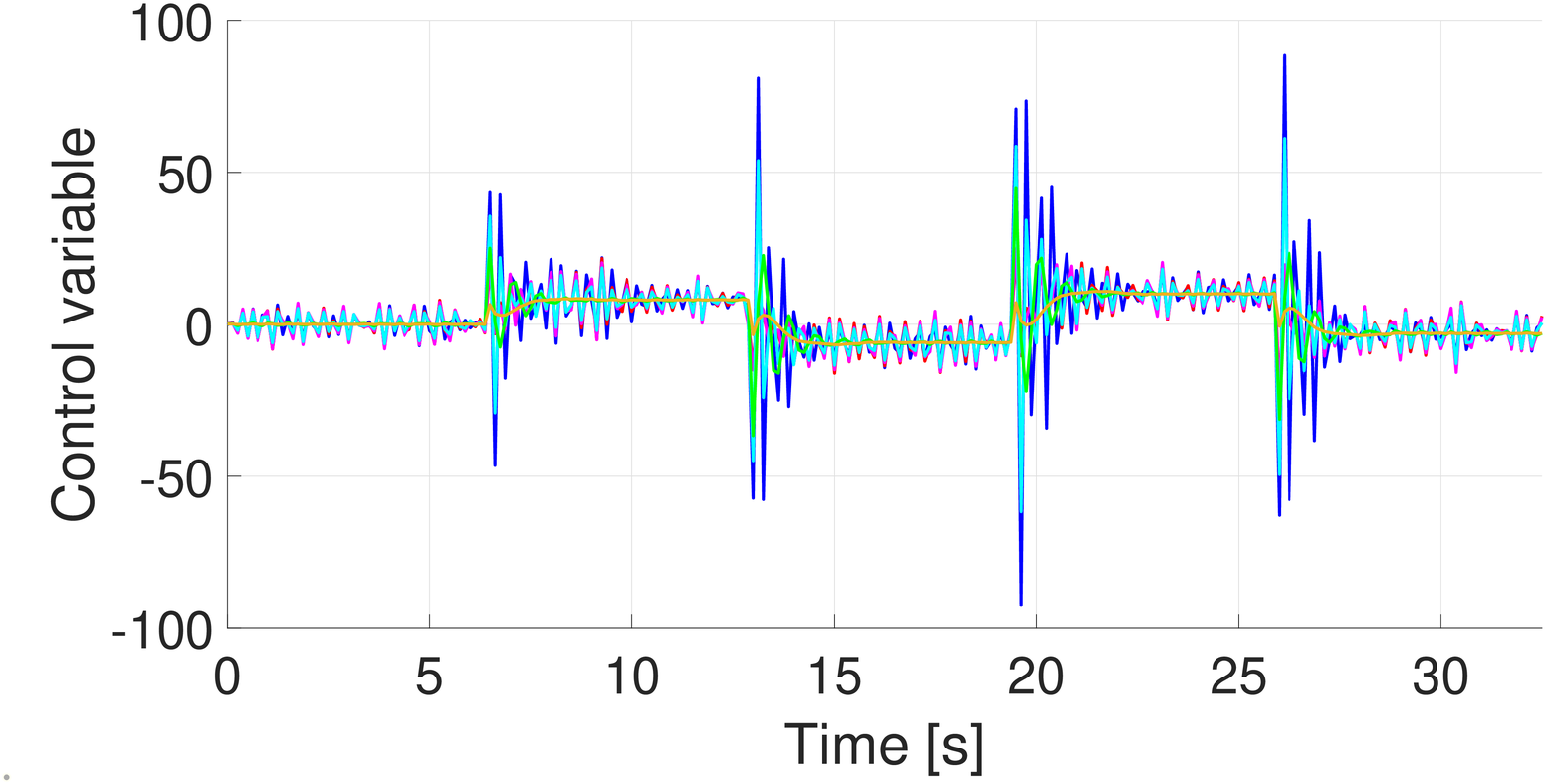}
	\caption{Input trajectories obtained with $\mathcal{M}_{10}$. Blue line: FF-SM-VRFT; cyan line: FF-SM-VRFT-NF; red line: EI-SM-VRFT; magenta line: EI-SM-VRFT-NF; golden line: PID-VRFT; green line: UF.}
	\label{cvar10}
\end{figure}
\begin{figure}[h!]
	\centering
	\includegraphics[width=0.75\columnwidth]{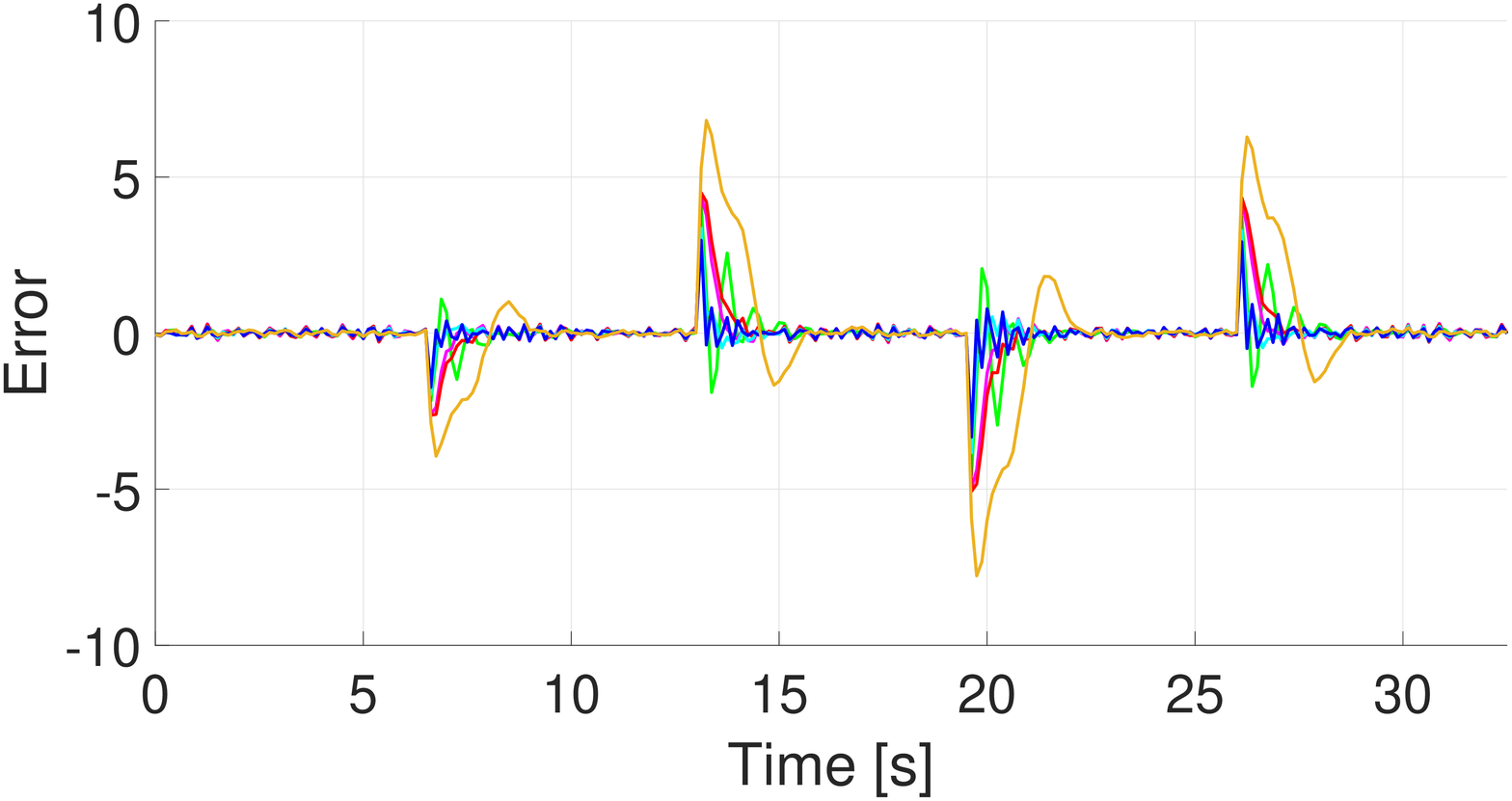}
	\caption{Error trajectories obtained with $\mathcal{M}_{10}$. Blue line: FF-SM-VRFT; cyan line: FF-SM-VRFT-NF; red line: EI-SM-VRFT; magenta line: EI-SM-VRFT-NF; golden line: PID-VRFT; green line: UF.}
	\label{err10}
\end{figure}
A similar trend is obtained also in the more challenging case where the reference model is $\mathcal{M}_{10}$, see Figures~\ref{out10},~\ref{cvar10}, and~\ref{err10}. In this case, being the settling time very short, the control system results to be extremely reactive, which is apparent from the output shown in Figure~\ref{out10}, and from the input variable trajectory shown in Figure~\ref{cvar10}. The best results are achieved in case of FF-SM-VRFT, while the PID tuned through the standard VRFT leads to unsatisfactory performances. Moreover, in this case, the application of the filter \eqref{filterapproxN} plays a marginal role, i.e., it does not improve the performances.
%

\subsection{Example 2: Non-minimum phase plant}
We consider a second example which shows the effectiveness of the proposed algorithms also in a more challenging scenario. In particular, a non-minimum phase system is considered, corresponding to the discretization of the asymptotically stable system with continuous-time transfer function:
\begin{align}
	\label{sysolnm}
	&P(s)=\frac{Z(s)}{U(s)}=\frac{160s-80}{(s+10)(s^2+1.6s+16)}&
\end{align}
characterized by a static gain $\mu=-0.5$, the same poles of \eqref{sysol}, and a real positive zero in $0.5$. A sample time $T_s=0.125$ s is chosen. The system is discretized by means of the zero-order hold (ZOH) method: the corresponding nominal parameter vector is $\theta^{o}=\begin{bmatrix}1.883&-1.276&0.2346&0.7617&-0.346&-0.4948\end{bmatrix}^T$ and the order is $n=3$. An additive uniform random noise $d$ acting in the range $[-0.1,0.1]$ (i.e., $\bar{d}=0.1$) affects the output $z$ of the system.\\
Also in this case, two datasets composed of $N_d=10000$ output/regressor data pairs $(y(k+1),\hat{\phi}(k))$ are collected from the plant in open-loop, with different noise realizations. The input signal is a PseudoRandom Binary Sequence (PRBS) in the range $[-10,10]$. \\
The same considerations and design choices made in Section \ref{mpp} for the application of Algorithm 1 hold also in this case. We obtained the estimate $\alpha^*_p=1.1007$. The corresponding set $\Theta(\alpha^*_p)$ has $998$ vertices. 
\begin{figure}[h!]
	\centering
	\includegraphics[width=0.75\columnwidth]{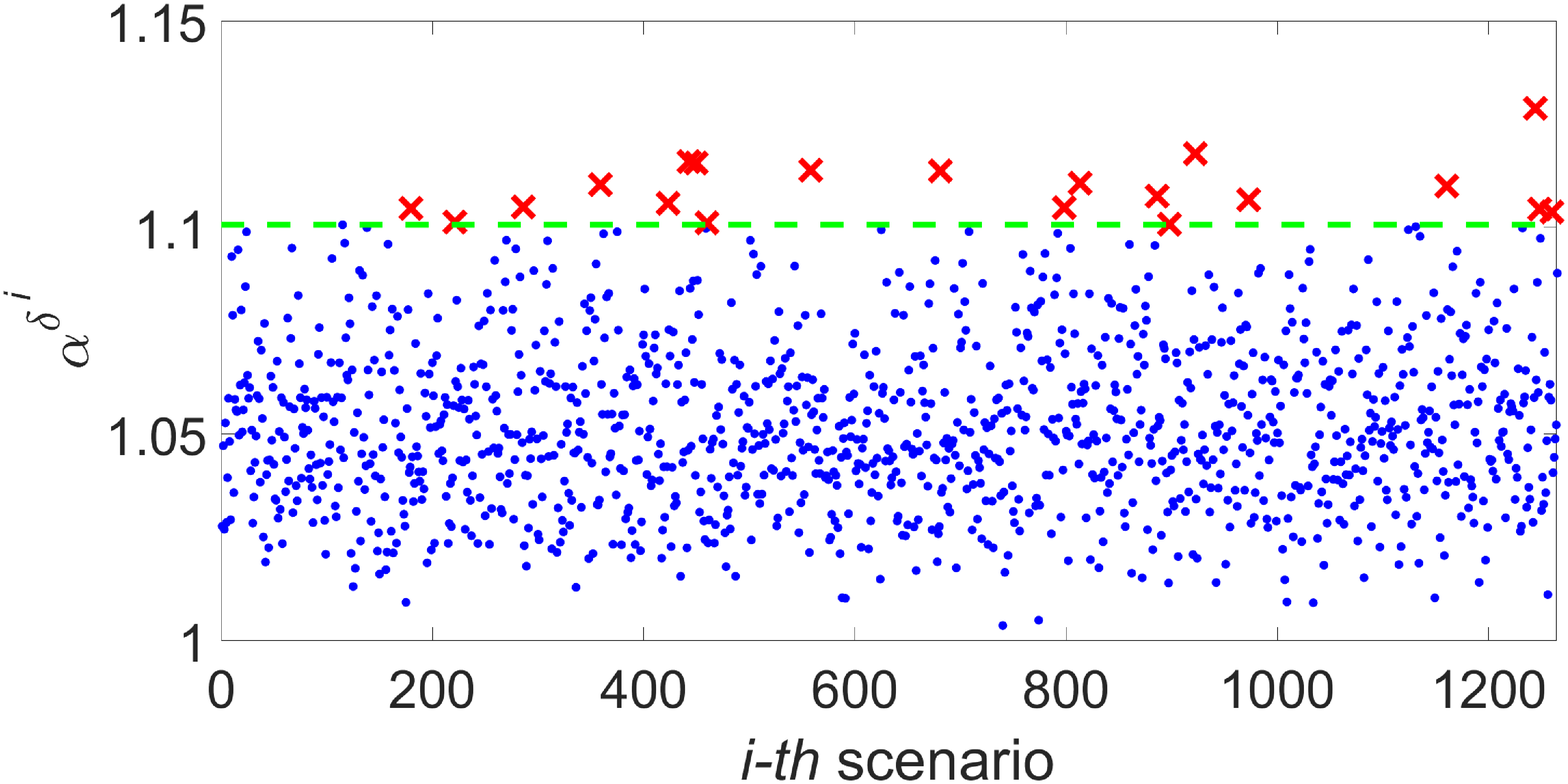}
	\caption{Scenario application: scenarios (blue dots), removed scenarios (red crosses), $\alpha^*_p$ value (green dashed line).}
	\label{sce1nm}
\end{figure}
In Figure \ref{sce1nm} the values $\alpha^{\delta^i}$ computed for each scenario $i$ are depicted. 
\begin{figure}[h!]
	\centering
	\includegraphics[width=0.75\columnwidth]{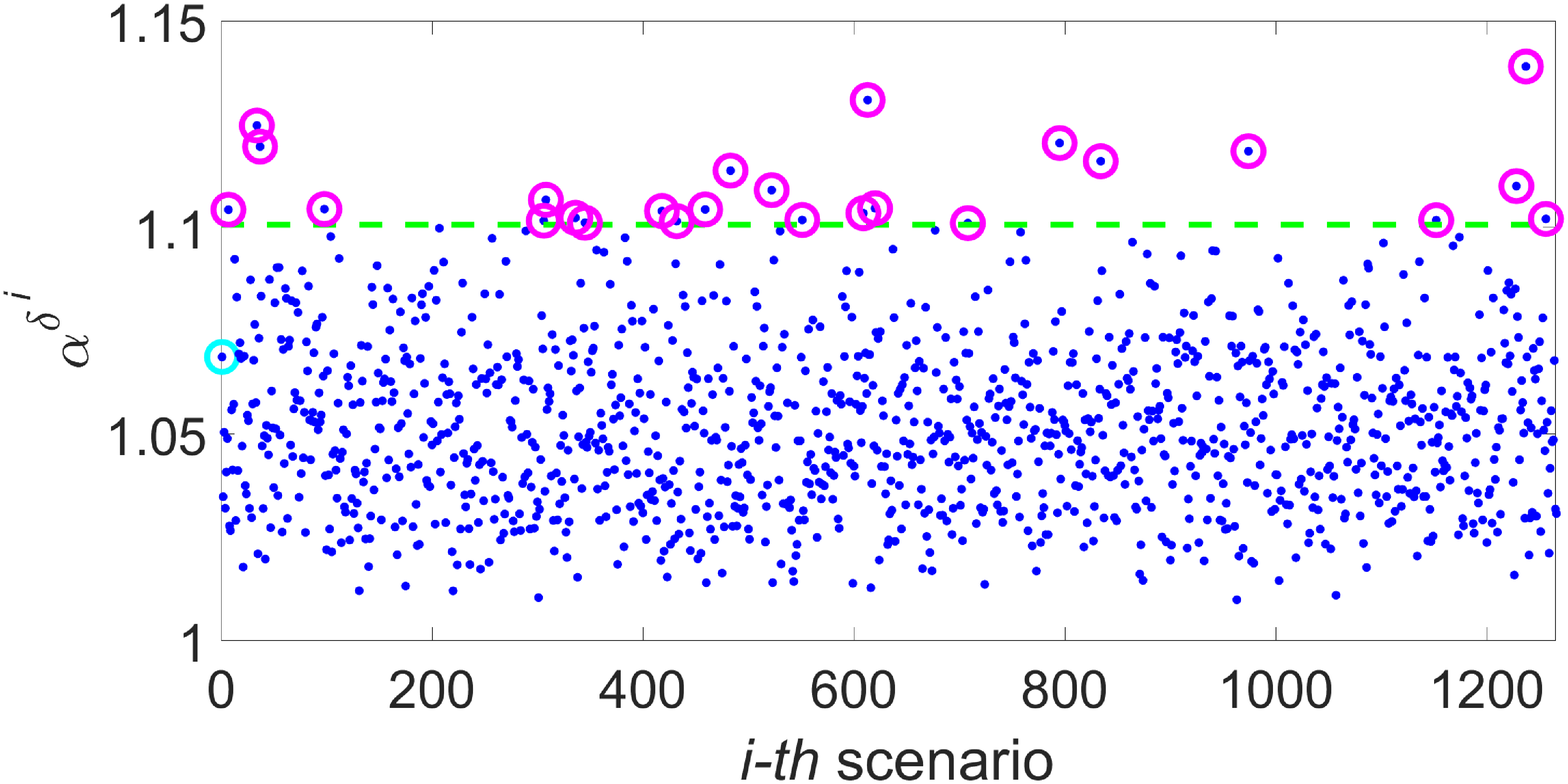}
	\caption{Validation: scenarios (blue dots), nominal case (cyan circle), violated scenarios (purple circles), $\alpha^*_p$ value (green dashed line).}
	\label{sce2nm}
\end{figure}
In Figure \ref{sce2nm} a validation test with new scenarios is carried out. The percentage of scenarios which violate $\alpha^*_p$ is $1.9763\%$, lower than $\epsilon_{\%}=5\%$. Note that the nominal parameter vector $\theta^o\in\Theta(\alpha^*_p)$. \\
\\
The following first-order asymptotically stable and unitary-gain system, denoted with $\mathcal{M}_{60}$, is selected as reference model: $y_r(k) = -a_1 y_r(k-1) + b_1 r(k-1)$, where $a_1=-0.925$, $b_1=0.075$, and the settling time is $60T_s=7.5$ s. This choice is not demanding in terms of settling time as in the previous example. However, since the inverse response of the system cannot be avoided due to the positive real part zero \eqref{sysolnm} and the reference model is a minimum phase system, the design turns out to be challenging. \\
\\
As for Example 1, a comparison between the performances achieved with the proposed Algorithms 2 and 3, the standard VRFT-based PID tuning, and the UF method is performed. The methods are denoted with the same notation used in Section \ref{mpp}. In the simulations, the reference signal described in Table~\ref{tab: Refsignval} is used (with longer time intervals) to evaluate the performances of the control system.\\
Again, we selected $c=10^6$ and $W(q)=1$, while $Z(q)$ was estimated from the output signal of one experiment through the identification of a discrete-time AR model of order 21. For the application of the UF method, the $\mathcal{M}_{60}$ complementary sensitivity function was used and the following suitable input sensitivity was considered: $$\mathcal{Q}(q)=\frac{-0.09012q^2+0.1439q-0.0738}{q^2-1.791q+0.801}$$
The class of controllers \eqref{ufcontr} is taken into account also in this case. In UF, the maximum value of the weight $\delta$ for which the stability test is passed is equal to $0.05$. \smallskip\\
\begin{table}[h!]
	\centering
	\renewcommand\arraystretch{1.2}
	\caption{Spectral radius of the closed-loop system and $FIT$} 
	\begin{tabular}{ c c c r | }
		\hline
		\multicolumn{1}{c}{$Case$} &
		\multicolumn{1}{c}{$\rho$} &
		\multicolumn{1}{c}{$FIT\ (\%) $} \\
		\hline
		FF-SM-VRFT $\mathcal{M}_{60}$ & $0.9757$ & $48.3004$ \\
		EI-SM-VRFT $\mathcal{M}_{60}$ & $0.9740$ & $45.1159$ \\		
		PID-VRFT $\mathcal{M}_{60}$ & $1.0066$ & $<0$ \\
		UF $\mathcal{M}_{60}$ & $0.9784$ & $57.2788$ \\		
		\hline
	\end{tabular} 
	\label{tab: spradnm} 
\end{table}
Table \ref{tab: spradnm} displays the spectral radius $\rho$ of the closed-loop system and the fitting index \eqref{fitsim}. 
\begin{figure}[h!]
	\centering
	\includegraphics[width=0.75\columnwidth]{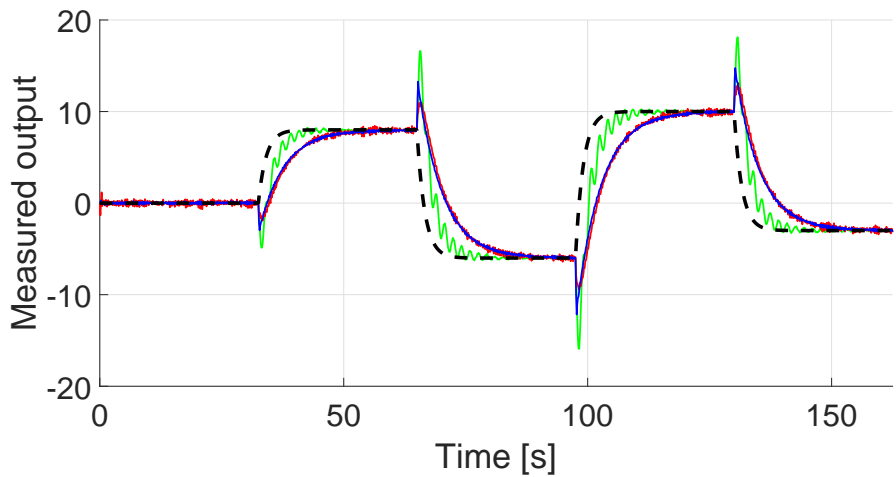}
	\caption{Measured output trajectories obtained with $\mathcal{M}_{60}$. Black dashed line: reference closed-loop trajectory; blue line: FF-SM-VRFT; red line: EI-SM-VRFT; green line: UF.}
	\label{out60}
\end{figure}
\begin{figure}[h!]
	\centering
	\includegraphics[width=0.75\columnwidth]{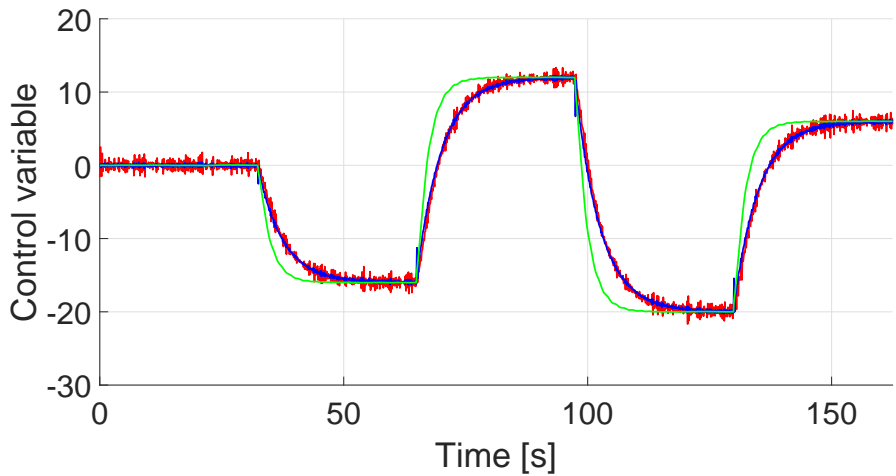}
	\caption{Input trajectories obtained with $\mathcal{M}_{60}$. Blue line: FF-SM-VRFT; red line: EI-SM-VRFT; green line: UF.}
	\label{cvar60}
\end{figure}
\begin{figure}[h!]
	\centering
	\includegraphics[width=0.75\columnwidth]{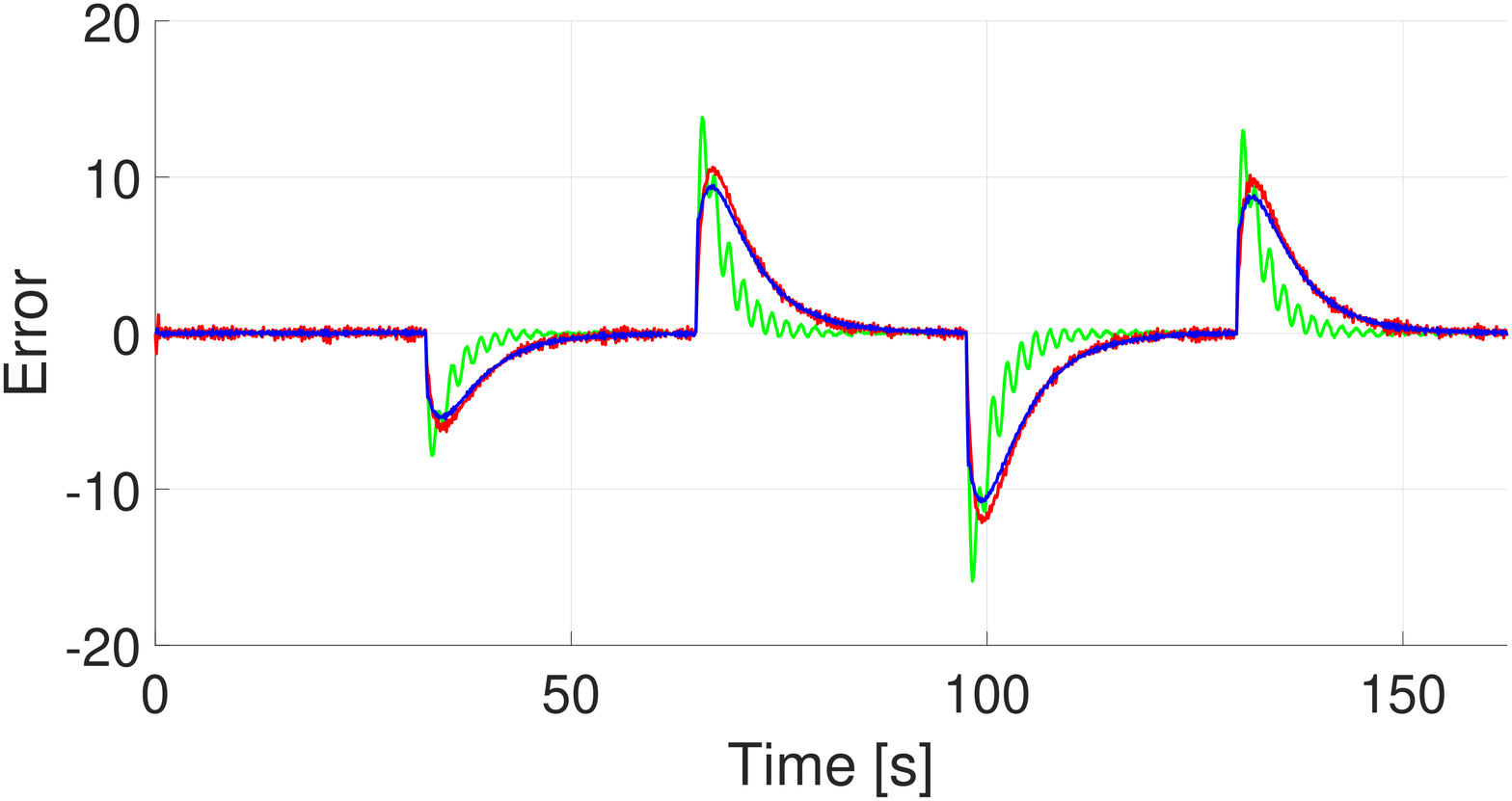}
	\caption{Error trajectories obtained with $\mathcal{M}_{60}$. Blue line: FF-SM-VRFT; cyan line: FF-SM-VRFT-NF; red line: EI-SM-VRFT; magenta line: EI-SM-VRFT-NF; green line: UF.}
	\label{err60}
\end{figure}
In Figures~\ref{out60} and~\ref{cvar60} we show the reference tracking results obtained with reference model $\mathcal{M}_{60}$, in terms of trajectories of the measured outputs $y(k)$ and of the control inputs $u(k)$, respectively. In Figure~\ref{err60} the output error trajectories are depicted. \\
Note that an unstable closed-loop system is obtained with the standard VRFT-based PID implementation, while asymptotically stable closed-loop systems are achieved with the other methods. Satisfactory tracking results follow from the application of the proposed Algorithms 2 and 3. However, the best results in terms of fitting are obtained with the UF method, even if, as evident from Figure \ref{out60}, its response is slightly underdamped. Furthermore, as specified, the UF method needs for an accurate design of a suitable reference control sensitivity function, requiring some knowledge about the model of the system \eqref{sysolnm}.
\section{Conclusions}
\label{sec:conclusions}
In this paper, a method for the application of VRFT with robust closed-loop stability guarantees based on a probabilistic Set Membership identification approach has been presented. The developed method involves only LMI optimization problems with linear cost functions and has been shown to be particularly effective for the design of controllers for tracking reference signals. Two simulation case studies have corroborated the effectiveness of the proposed algorithm showing the potentialities of the method and significant advantages with respect to the classical VRFT.\\
Future works include the possibility of integrating the controllers, designed using the proposed method, in the MPC-based scheme discussed in \cite{piga2017direct}, in order to cope with constraints on input and output variables. Also, the use of regularization methods \cite{ZecevicSiljac} is envisaged, to reduce the control effort. \\
A further interesting follow-up theoretical research consists of the application of the method described here to the nonlinear case, considering notable classes of nonlinear systems (e.g., recurrent neural networks). Finally, the proposed approach will be used in experimental cases to further validate its effectiveness.
\section*{Acknowledgements}
The authors would like to thank Simone Garatti and Giulio Panzani for the useful suggestions and the fruitful discussions.

\bibliographystyle{ieeetr}
\bibliography{Bibliografia.bib}

\begin{thebibliography}{10}

\bibitem{hou2013model}
Z.-S. Hou and Z.~Wang, ``From model-based control to data-driven control:
  Survey, classification and perspective,'' {\em Information Sciences},
  vol.~235, pp.~3--35, 2013.

\bibitem{armenio2019model}
L.~B. Armenio, E.~Terzi, M.~Farina, and R.~Scattolini, ``Model predictive
  control design for dynamical systems learned by echo state networks,'' {\em
  IEEE Control Systems Letters}, vol.~3, no.~4, pp.~1044--1049, 2019.

\bibitem{campi2000virtual}
M.~Campi, A.~Lecchini, and S.~M. Savaresi, ``Virtual reference feedback tuning
  ({VRFT}): a new direct approach to the design of feedback controllers,'' in
  {\em Proceedings of the 39th IEEE Conference on Decision and Control (Cat.
  No. 00CH37187)}, vol.~1, pp.~623--629, IEEE, 2000.

\bibitem{hjalmarsson2005experiment}
H.~Hjalmarsson, ``From experiment design to closed-loop control,'' {\em
  Automatica}, vol.~41, no.~3, pp.~393--438, 2005.

\bibitem{feldbaum1963dual}
{\'{}A}.~Feldb{\^a}um, ``Dual control theory problems,'' {\em IFAC Proceedings
  Volumes}, vol.~1, no.~2, pp.~541--550, 1963.

\bibitem{formentin2018core}
S.~Formentin and A.~Chiuso, ``Core: Control-oriented regularization for system
  identification,'' in {\em 2018 IEEE Conference on Decision and Control
  (CDC)}, pp.~2253--2258, IEEE, 2018.

\bibitem{dorfler2022bridging}
F.~Dorfler, J.~Coulson, and I.~Markovsky, ``Bridging direct \& indirect
  data-driven control formulations via regularizations and relaxations,'' {\em
  IEEE Transactions on Automatic Control}, 2022.

\bibitem{dehghani2009validating}
A.~Dehghani, A.~Lecchini-Visintini, A.~Lanzon, and B.~D. Anderson, ``Validating
  controllers for internal stability utilizing closed-loop data,'' {\em IEEE
  Transactions on Automatic Control}, vol.~54, no.~11, pp.~2719--2725, 2009.

\bibitem{cha2014verifying}
S.~H. Cha, A.~Dehghani, W.~Chen, and B.~D. Anderson, ``Verifying stabilizing
  controllers for performance improvement using closed-loop data,'' {\em
  International Journal of Adaptive Control and Signal Processing}, vol.~28,
  no.~2, pp.~121--137, 2014.

\bibitem{da2020one}
G.~R.~G. da~Silva, A.~S. Bazanella, and L.~Campestrini, ``One-shot data-driven
  controller certification,'' {\em ISA transactions}, vol.~99, pp.~361--373,
  2020.

\bibitem{van2011data}
K.~Van~Heusden, A.~Karimi, and D.~Bonvin, ``Data-driven model reference control
  with asymptotically guaranteed stability,'' {\em International Journal of
  Adaptive Control and Signal Processing}, vol.~25, no.~4, pp.~331--351, 2011.

\bibitem{battistelli2018direct}
G.~Battistelli, D.~Mari, D.~Selvi, and P.~Tesi, ``Direct control design via
  controller unfalsification,'' {\em International Journal of Robust and
  Nonlinear Control}, vol.~28, no.~12, pp.~3694--3712, 2018.

\bibitem{selvi2021optimal}
D.~Selvi, D.~Piga, G.~Battistelli, and A.~Bemporad, ``Optimal direct
  data-driven control with stability guarantees,'' {\em European Journal of
  Control}, vol.~59, pp.~175--187, 2021.

\bibitem{de2019formulas}
C.~De~Persis and P.~Tesi, ``Formulas for data-driven control: Stabilization,
  optimality, and robustness,'' {\em IEEE Transactions on Automatic Control},
  vol.~65, no.~3, pp.~909--924, 2019.

\bibitem{dorfler2021certainty}
F.~D{\"o}rfler, P.~Tesi, and C.~De~Persis, ``On the certainty-equivalence
  approach to direct data-driven lqr design,'' {\em arXiv preprint
  arXiv:2109.06643}, 2021.

\bibitem{coulson2021distributionally}
J.~Coulson, J.~Lygeros, and F.~Dorfler, ``Distributionally robust chance
  constrained data-enabled predictive control,'' {\em IEEE Transactions on
  Automatic Control}, 2021.

\bibitem{breschi2021direct}
V.~Breschi, C.~D. Persis, S.~Formentin, and P.~Tesi, ``Direct data-driven
  model-reference control with lyapunov stability guarantees,'' in {\em 2021
  60th IEEE Conference on Decision and Control (CDC)}, pp.~1456--1461, 2021.

\bibitem{bisoffi2021trade}
A.~Bisoffi, C.~De~Persis, and P.~Tesi, ``Trade-offs in learning controllers
  from noisy data,'' {\em Systems \& Control Letters}, vol.~154, p.~104985,
  2021.

\bibitem{de2021low}
C.~De~Persis and P.~Tesi, ``Low-complexity learning of linear quadratic
  regulators from noisy data,'' {\em Automatica}, vol.~128, p.~109548, 2021.

\bibitem{novara2014set}
C.~Novara, M.~Canale, M.~Milanese, and M.~Signorile, ``Set membership inversion
  and robust control from data of nonlinear systems,'' {\em International
  Journal of Robust and Nonlinear Control}, vol.~24, no.~18, pp.~3170--3195,
  2014.

\bibitem{guo2021data}
M.~Guo, C.~De~Persis, and P.~Tesi, ``Data-driven stabilization of nonlinear
  polynomial systems with noisy data,'' {\em IEEE Transactions on Automatic
  Control}, 2021.

\bibitem{campi2002virtual}
M.~C. Campi, A.~Lecchini, and S.~M. Savaresi, ``Virtual reference feedback
  tuning: a direct method for the design of feedback controllers,'' {\em
  Automatica}, vol.~38, no.~8, pp.~1337--1346, 2002.

\bibitem{campi2006direct}
M.~C. Campi and S.~M. Savaresi, ``Direct nonlinear control design: The virtual
  reference feedback tuning ({VRFT}) approach,'' {\em IEEE Transactions on
  Automatic Control}, vol.~51, no.~1, pp.~14--27, 2006.

\bibitem{sala2005extensions}
A.~Sala and A.~Esparza, ``Extensions to “virtual reference feedback tuning: A
  direct method for the design of feedback controllers”,'' {\em Automatica},
  vol.~41, no.~8, pp.~1473--1476, 2005.

\bibitem{sala2005virtual}
A.~Sala and A.~Esparza, ``Virtual reference feedback tuning in restricted
  complexity controller design of non-minimum phase systems,'' {\em IFAC
  Proceedings Volumes}, vol.~38, no.~1, pp.~235--240, 2005.

\bibitem{rojas2011internal}
J.~D. Rojas and R.~Vilanova, ``Internal model controller tuning using the
  virtual reference approach with robust stability,'' {\em IFAC Proceedings
  Volumes}, vol.~44, no.~1, pp.~10237--10242, 2011.

\bibitem{chiluka2021novel}
S.~K. Chiluka, S.~R. Ambati, M.~M. Seepana, and U.~B.~B. Gara, ``A novel robust
  virtual reference feedback tuning approach for minimum and non-minimum phase
  systems,'' {\em ISA transactions}, vol.~115, pp.~163--191, 2021.

\bibitem{milanese2013bounding}
M.~Milanese, J.~Norton, H.~Piet-Lahanier, and {\'E}.~Walter, {\em Bounding
  approaches to system identification}.
\newblock Springer Science \& Business Media, 2013.

\bibitem{terzi2019learning}
E.~Terzi, L.~Fagiano, M.~Farina, and R.~Scattolini, ``Learning-based predictive
  control for linear systems: A unitary approach,'' {\em Automatica}, vol.~108,
  p.~108473, 2019.

\bibitem{lauricella2020set}
M.~Lauricella and L.~Fagiano, ``Set membership identification of linear systems
  with guaranteed simulation accuracy,'' {\em IEEE Transactions on Automatic
  Control}, vol.~65, no.~12, pp.~5189--5204, 2020.

\bibitem{campi2011sampling}
M.~C. Campi and S.~Garatti, ``A sampling-and-discarding approach to
  chance-constrained optimization: feasibility and optimality,'' {\em Journal
  of optimization theory and applications}, vol.~148, no.~2, pp.~257--280,
  2011.

\bibitem{de1999new}
M.~C. De~Oliveira, J.~Bernussou, and J.~C. Geromel, ``A new discrete-time
  robust stability condition,'' {\em Systems \& control letters}, vol.~37,
  no.~4, pp.~261--265, 1999.

\bibitem{boyd1993control}
S.~Boyd, V.~Balakrishnan, E.~Feron, and L.~ElGhaoui, ``Control system analysis
  and synthesis via linear matrix inequalities,'' in {\em 1993 American Control
  Conference}, pp.~2147--2154, IEEE, 1993.

\bibitem{kothare1996robust}
M.~V. Kothare, V.~Balakrishnan, and M.~Morari, ``Robust constrained model
  predictive control using linear matrix inequalities,'' {\em Automatica},
  vol.~32, no.~10, pp.~1361--1379, 1996.

\bibitem{cerone2017direct}
V.~Cerone, D.~Regruto, and M.~Abuabiah, ``Direct data-driven control design
  through set-membership errors-in-variables identification techniques,'' in
  {\em 2017 American Control Conference (ACC)}, pp.~388--393, IEEE, 2017.

\bibitem{cerone2017set}
V.~Cerone, D.~Regruto, and M.~Abuabiah, ``A set-membership approach to direct
  data-driven control design for siso non-minimum phase plants,'' in {\em 2017
  IEEE 56th Annual Conference on Decision and Control (CDC)}, pp.~1284--1290,
  IEEE, 2017.

\bibitem{van2012subspace}
P.~Van~Overschee and B.~De~Moor, {\em Subspace identification for linear
  systems: Theory—Implementation—Applications}.
\newblock Springer Science \& Business Media, 2012.

\bibitem{avis2009polyhedral}
D.~Avis, D.~Bremner, and A.~Deza, {\em Polyhedral computation}, vol.~48.
\newblock American Mathematical Soc., 2009.

\bibitem{bemporad2004inner}
A.~Bemporad, C.~Filippi, and F.~D. Torrisi, ``Inner and outer approximations of
  polytopes using boxes,'' {\em Computational Geometry}, vol.~27, no.~2,
  pp.~151--178, 2004.

\bibitem{ljung1998system}
L.~Ljung, ``System identification,'' in {\em Signal analysis and prediction},
  pp.~163--173, Springer, 1998.

\bibitem{lofberg2004yalmip}
J.~Lofberg, ``Yalmip: A toolbox for modeling and optimization in matlab,'' in
  {\em 2004 IEEE international conference on robotics and automation (IEEE Cat.
  No. 04CH37508)}, pp.~284--289, IEEE, 2004.

\bibitem{mosek}
M.~ApS, {\em The MOSEK optimization toolbox for MATLAB manual. Version 9.0},
  2019.

\bibitem{care2019toolbox}
A.~Car{\`e}, F.~Torricelli, M.~C. Campi, and S.~M. Savaresi, ``A toolbox for
  virtual reference feedback tuning (vrft),'' in {\em 2019 18th European
  control conference (ECC)}, pp.~4252--4257, IEEE, 2019.

\bibitem{piga2017direct}
D.~Piga, S.~Formentin, and A.~Bemporad, ``Direct data-driven control of
  constrained systems,'' {\em IEEE Transactions on Control Systems Technology},
  vol.~26, no.~4, pp.~1422--1429, 2017.

\bibitem{ZecevicSiljac}
A.~Zecevic and D.~Siljac, {\em Control of Complex Systems. Structural
  Constraints and Uncertainty}.
\newblock Springer, 2010.

\end{thebibliography}

\end{document}